\documentclass[11pt,onecolumn,draftcls,journal]{IEEEtran}
  \usepackage[pdftex]{graphicx}
  \usepackage{enumerate}
  \usepackage{cite}
  \graphicspath{{./images/}}
  \DeclareGraphicsExtensions{.pdf,.png}
\usepackage[cmex10]{amsmath}

\DeclareMathOperator*{\bigtimes}{ \textsf{X}}

\usepackage[caption=false,font=footnotesize]{subfig}
\usepackage{fixltx2e}
\usepackage{amssymb,bm}	
\usepackage{epstopdf}		

\usepackage{etoolbox}

\newtoggle{comment}

\togglefalse{comment}   

\newcommand{\edits}{\iftoggle{comment}{\textbf{\textit{(editing here) }}}{}}

\def\R{ \mathbb{R}}
\def\t{ {\bf x}}
\def\t{ {\bf t}}
\def\s{{\bf s}}

\def\n{ \mathbb{n}}

\def\M{ \mathbb{M}}

\def\M{ \mathcal{M}}
\def\Ncal{\mathcal{N}}

\def\B{ \mathcal{B}}
\def\u{ {\bf u}}

\def\n{ {\bf n} }
\def\v{ {\bf v}}

\def\k{ {\bf k}}

\def\A{ \mathcal{A}}
\def\O{ \mathcal{O}}

\def\rect{ \text{rect} }
\def\sinc{ \text{sinc} }

\def\DFT{ \text{DFT} }
\def\IDFT{ \text{IDFT} }

\def\bbeta{ {\bm \beta}}

\newtheorem{definition}{Definition}
\newtheorem{theorem}{Theorem}
\newtheorem{lemma}{Lemma}
\newtheorem{fact}{Fact}
\newtheorem{corollary}{Corollary}

\newcommand{\tvec}{\mathbf t}
\newcommand{\svec}{\mathbf s}
\newcommand{\nvec}{\mathbf n}

\newcommand{\kvec}{\mathbf k}
\newcommand{\evec}{\mathbf e}
\newcommand{\lambdavec}{\bm \lambda}
\newcommand{\alphavec}{\bm \alpha}

\newcommand{\ovec}{\bm 0}
\newcommand{\Btilde}{\widetilde{\mathcal B}}
\newcommand{\Ntilde}{\widetilde{\mathcal N}}
\newcommand{\Mtilde}{\widetilde{\mathcal M}}

\newcommand{\bvec}{\mathbf b}
\newcommand{\onevec}{\mathbf 1}
\newcommand{\zerovec}{\mathbf 0}

\begin{document}
%
\title{Multidimensional Manhattan Sampling\\  and Reconstruction}
\author{Matthew A. Prelee and 
        David L. Neuhoff \\[2ex]
%
%
\thanks{The authors are with the Department of Electrical Engineering and Computer Science, University of Michigan, Ann Arbor, MI 48109 ~(email:  mprelee@umich.edu; neuhoff@eecs.umich.edu).}
\thanks{This work was supported by NSF Grant CCF 0830438.
Portions were presented at IEEE ICASSP 2012
and  ITA 2014. \edits}
}

\maketitle

\begin{abstract} 
  %
  This paper introduces Manhattan sampling in two and higher dimensions, and
  proves sampling theorems.  In two dimensions, Manhattan sampling, which
  takes samples densely along a Manhattan grid of lines, can be viewed as
  sampling on the union of two rectangular lattices,  one dense
  horizontally, 
  the other vertically,
  with the coarse spacing of each 
  being a multiple of the fine spacing of the other.  
  The sampling theorem shows that images
  bandlimited to the union of the Nyquist regions of the two rectangular
  lattices can be recovered from their Manhattan samples, and an efficient
  procedure for doing so is given.  Such recovery is possible even though there is
  overlap among the spectral replicas induced by Manhattan
  sampling.
  \iftoggle{comment}{\edits}{}
  
  In three and higher dimensions, there are many possible configurations for 
  Manhattan sampling, each consisting of the union of
  special rectangular lattices called bi-step lattices.
  This paper identifies them, proves a sampling theorem
  showing that images bandlimited to the union of the Nyquist regions of the
  bi-step rectangular lattices are recoverable from Manhattan samples, and presents
  an efficient onion-peeling procedure for doing so.
  Furthermore, it develops a special representation for the bi-step lattices
  and an algebra with nice properties.
  It is also shown that the set of reconstructable images is
  maximal in the Landau sense.  
  
  While  most of the paper deals with continuous-space images, Manhattan
  sampling of discrete-space images is also considered, for infinite, as well as finite, support images.
\end{abstract}



%
\IEEEpeerreviewmaketitle

\section{Introduction}
%
%
%
%

\begin{figure}[!t] \centerline{ 
  \subfloat[]{\includegraphics[width=1.77in]{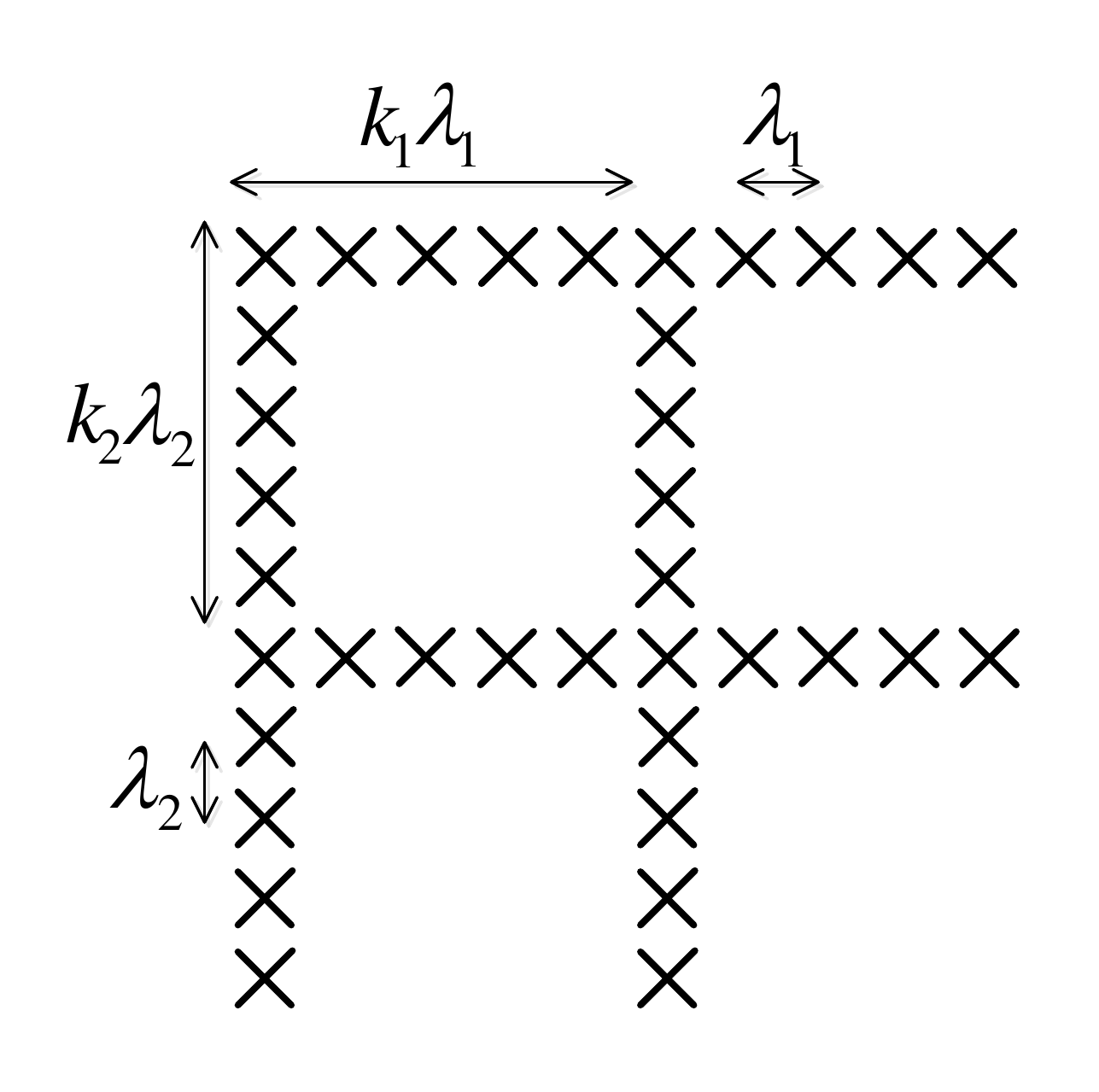}
  %
  }
 \hspace{.005in} 
 \subfloat[]{\includegraphics[width=1.55in]{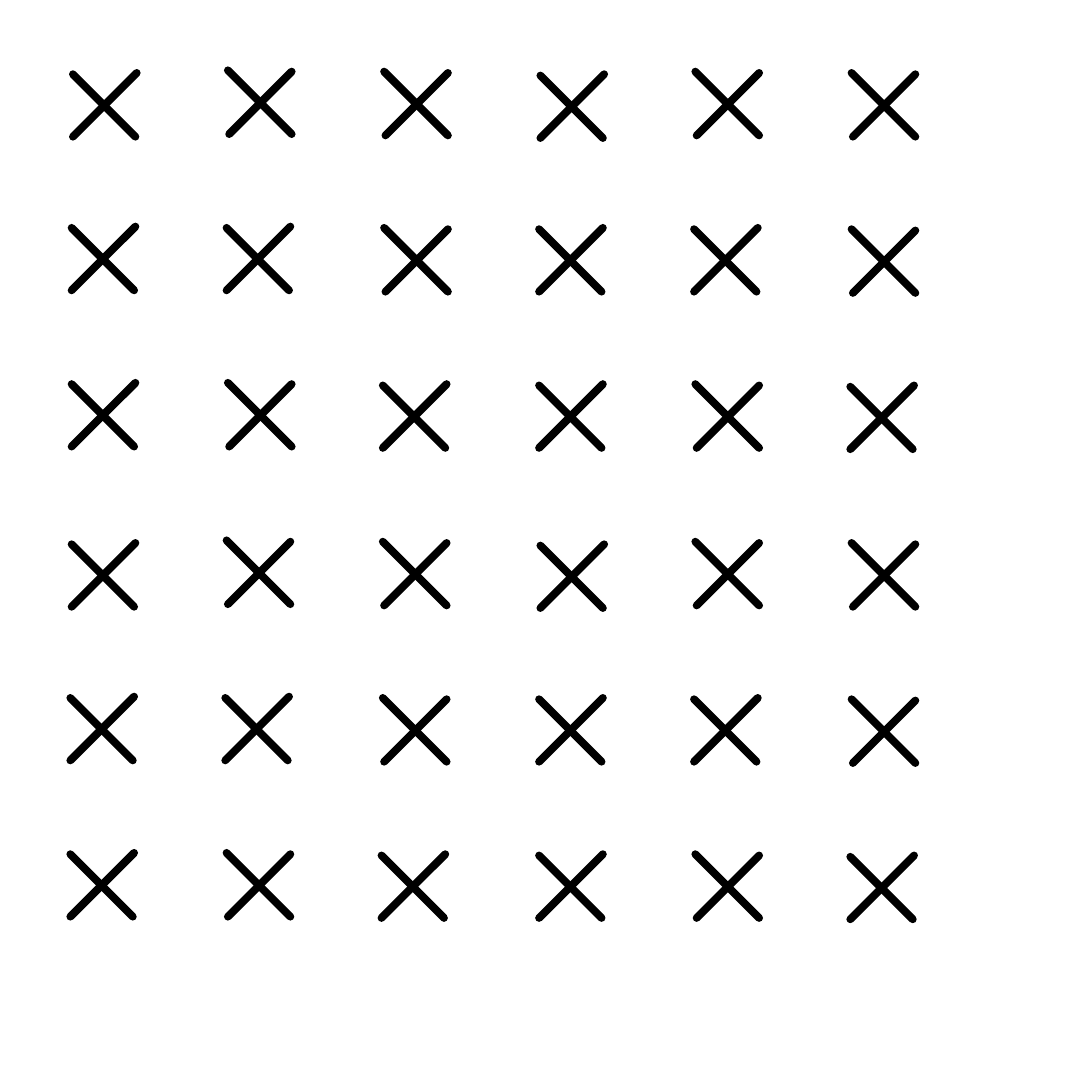}
  %
  }
  %
  %
  \subfloat[]{\includegraphics[width=1.55in]{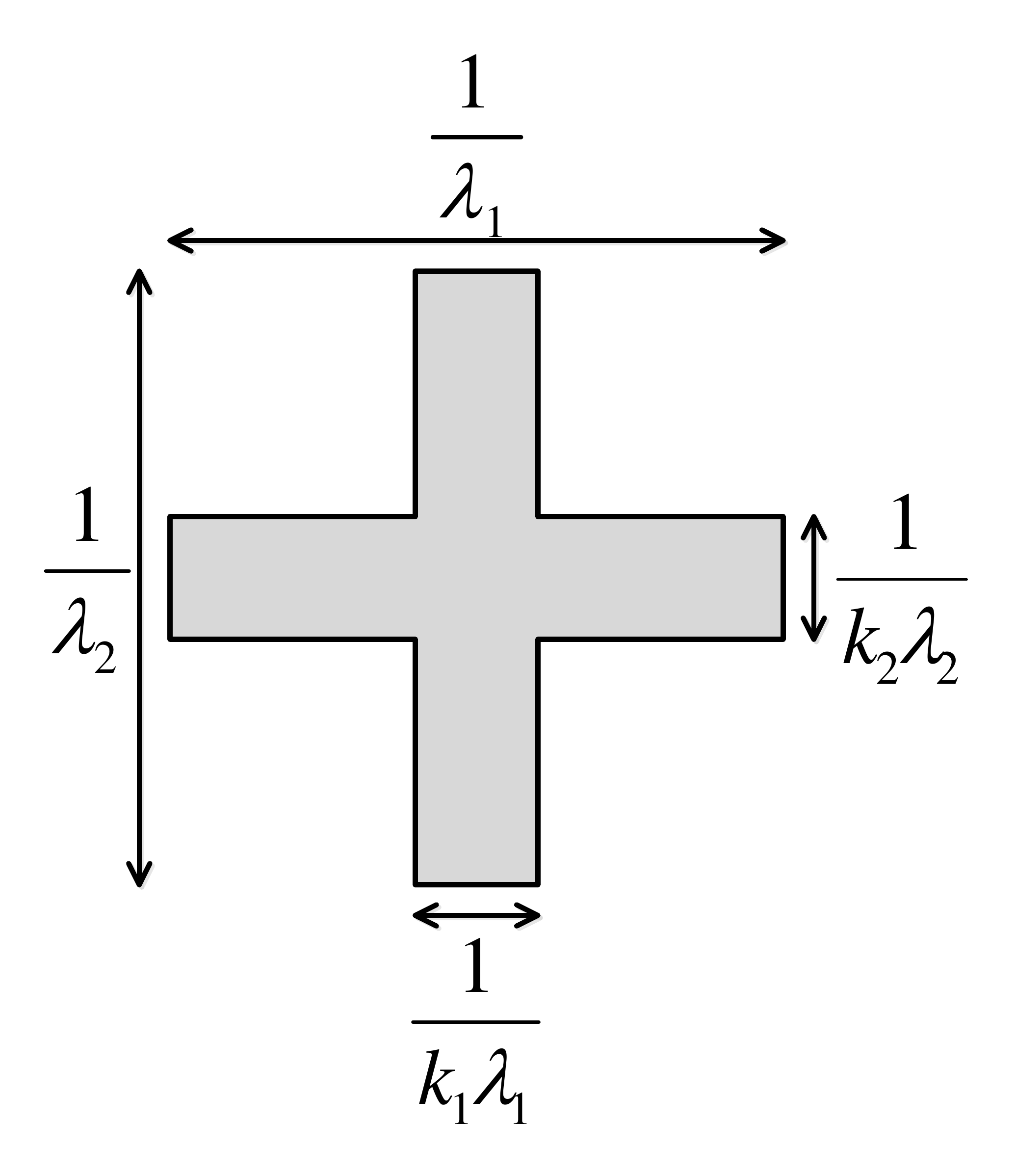}
  %
  }
  %
  %
  %
  %
  } 
  \caption{ (a)  2D Manhattan-grid sampling sites with parameters $k_1=k_2=5$
  and $\lambda_1=\lambda_2$.  (b) Square lattice sampling at the
  same density.  (c) Cross-shaped frequency support (centered at the origin) of images recoverable with
  Manhattan sampling. 
  }
  \label{fig:twodim} 
\end{figure}

\IEEEPARstart{I}{n} the two-dimensional (2D) setting, \emph{Manhattan sampling} (or
\emph{M-sampling} for short) is a recently proposed form of image sampling in which data is
taken along evenly spaced rows and columns; the set of sample locations will be
called a \emph{Manhattan grid}.
  In particular, as illustrated in Fig.\,\ref{fig:twodim}(a), given sampling intervals $\lambda_1 , \lambda_2 >0$ and
integers $k_1, k_2 > 1$,  samples are taken at intervals of $\lambda_1$
along horizontal rows spaced $k_2\lambda_2$ apart, and also at intervals of
$\lambda_2$ along vertical columns spaced $k_1\lambda_1$ apart.  

Manhattan sampling has been used to good effect in
both lossy and lossless bilevel image compression
\cite{reyes07lossy,reyes09arithmetic,reyes10lossless}.  These methods 
losslessly compress the samples in a Manhattan grid, for example with arithmetic
coding (AC), 
which can be done with very few bits per M-sample because
the samples are closely spaced and, hence, highly correlated.  
For lossless compression, the other pixels are then
AC encoded, conditioned on those in the Manhattan grid, while for lossy compression
there is no further encoding, and the decoder estimates the remaining pixels
from those in the Manhattan grid.  Markov random field models have been used to
guide both the arithmetic coding and the estimation.   

M-sampling has also been proposed
\cite{farmer2011cutset,prelee2012image, prelee:icassp13, prelee:icassp14} as a new approach to sampling grayscale
images and other two-dimensional fields, 
with the motivations that (a) dense sampling along lines might capture
edge transitions more completely than conventional lattice sampling with the
same density, (b) sensor networks with a Manhattan deployment geometry need
less power or less wire to transmit data than conventional lattice or random deployments
at the same density \cite{prelee:icassp13,prelee:icassp14}, 
and (c) there are physical scenarios for which M-sampling 
is far more natural than traditional lattice sampling, such as when
sampling from a moving vehicle, e.g., a ship sampling
oxygen levels in a body of water.  Similarly motivated by sampling
from vehicles, the recent related work of Unnikrishnan and Vetterli \cite{unnikrishnan:IT13,unnikrishnan:SP13}
considers sampling continuously along a grid of lines, 
i.e., with asymptotically large sampling rate.  

Methods for approximately reconstructing typical (non-bandlimited) images from M-samples have been 
developed in \cite{farmer2011cutset, prelee2012image, prelee2014icip}.  The present paper focuses on 
identifying a bandlimited set of images that can be perfectly reconstructed, as well as
efficient methods for doing so.


Manhattan sampling with parameters $\lambda_1, \lambda_2, k_1, k_2$  can be
viewed as sampling on the union of the horizontally dense rectangular lattice
consisting of all locations of the form $(n_1 \lambda_1, n_2 k_2 \lambda_2)$, where
$n_1, n_2$ are arbitrary integers, and the similarly defined vertically dense
rectangular lattice consisting of all locations of the form $(n_1 k_1 \lambda_1,
n_2 \lambda_2)$.  For brevity, we call these the \emph{horizontal}
and \emph{vertical} lattices, respectively.

By the conventional 2D sampling theorem
\cite{peterson62sampling} (see also \cite[p.\,72]{rosenfeld76},\cite[Chap.\,3]{tekalp95digital},\cite[p.\,43]{dudgeon84multidimensional}), the samples on the horizontal
lattice are sufficient to distinguish and reconstruct any image
bandlimited to the Nyquist region $\big \{ (u,v) : |u| < { 1 \over 2
\lambda_1},   |v| < { 1 \over 2 k_2 \lambda_2} \big \}$. 
Likewise the
samples on the vertical lattice are sufficient to
distinguish and reconstruct any image bandlimited to the Nyquist region $\big
\{ (u,v) : |u| < { 1 \over 2 k_1 \lambda_1},   |v| < { 1 \over 2
\lambda_2} \}$.  
Each of these samplings is maximally efficient in the Landau
sense \cite{landau67necessary} that their sampling densities are as small as
the area of the Nyquist region.
Equivalently, the set of images bandlimited
to the Nyquist region is maximal for the given sampling scheme.

The first result of the present paper is a sampling theorem in Section
\ref{sec:twoD} showing that images bandlimited to the union of these two
Nyquist regions can be reconstructed from their samples on the union of the two
rectangular lattices, i.e., on the Manhattan grid, 
and an efficient procedure for doing so is given.  It is also shown that the
images bandlimited in this way form a maximal reconstructable set for the
Manhattan grid samples. 
As illustrated in Fig.\,\ref{fig:twodim}(c), the union of the two Nyquist
regions is the cross-shaped \emph{Manhattan region}.   We say that images whose
spectra are confined to such a region are \emph{Manhattan-bandlimited}. 
Given the relevance of Manhattan-bandlimiting, a figure in Section \ref{sec:twoD} 
will display
the effect of several instances of such on a typical image.
\edits


%




The principal goals of the remainder of the paper are to formulate
M-sampling in three and higher dimensions, and to derive a sampling
theorem and a  reconstruction procedure.
%
%
%
M-sampling in three dimensions can be motivated by the need to spatially sample
a three-dimensional volume with a vehicle, or to spatio-temporally sample a two-dimensional
region, as in video, or a spatio-temporal sensor network.  
Four-dimensional sampling can be motivated by the need
for spatio-temporal sampling of a three-dimensional spatial region.

In three and higher dimensions, M-sampling can take a variety of
forms.  In order to describe two of these in three dimensions, consider the
partition of 3D space into  $k_1 \lambda_1 \times k_2 \lambda_2
\times k_3 \lambda_3$ orthotopes (3D rectangles).   As
illustrated in Fig.\,\ref{fig:3d_examples}(a), one form of M-sampling
takes samples uniformly along each edge of each of these orthotopes --- with
spacing $\lambda_i$  along edges parallel to axis $i$.  Another form (Fig.\,\ref{fig:3d_examples}(c)) takes samples uniformly on each face of each
orthotope ---  with the samples on the face orthogonal to axis $i$ taken
according to a $\lambda_j \times \lambda_k$ rectangular lattice, where $j$ and
$k$ denote the other dimensions.  In other words, the first form samples
densely along lines and the second samples densely along hyperplanes.
Neither of these takes samples in the interior of any of the
aforementioned orthotopes.

More generally, as described in Section \ref{sec:higherD}, M-sampling
in an arbitrary dimension $d$ is defined as taking samples on the
union of some collection
of $d$-dimensional \emph{bi-step} lattices, which are rectangular
lattices defined by step sizes that in
dimension $i$ are restricted to  $\lambda_i$ or $k_i \lambda_i$.  Thus, there
are many possible M-samplings in $d$ dimensions, even when
$\lambda_i$'s and $k_i$'s are fixed.
We call such unions of $d$-dimensional bi-step lattices 
\emph{Manhattan sets}.


The main results of Sec.\,\ref{sec:higherD} are (a) a sampling theorem showing that
images bandlimited to the union of the Nyquist regions of the $d$-dimensional bi-step
lattices comprising the Manhattan set
can be distinguished by
their M-samples, (b) efficient, onion-peeling procedures for
perfectly reconstructing $d$-dimensional images, bandlimited as in (a), 
from their M-samples (one in frequency domain and one in spatial domain),
and (c) a proof that the set of such bandlimited
images is maximal in the Landau sense.


The development of the sampling theorem and reconstruction procedures
are enabled by an efficient 
parametrization of a bi-step lattice (with a given set of $\lambda_i$'s
and $k_i$'s) by a binary vector $\bvec = (b_1, \ldots, b_d)$ indicating 
the dimensions $i$ along which the
spacing between lattice points is the smaller value, $\lambda_i$, rather than
the larger value,  $k_i \lambda_i$.
This enables any Manhattan set to be compactly described by 
a finite set of $\bvec_i$'s (in addition to the $\lambda_i$'s
and $k_i$'s).   
A number of properties and relationships are enabled by this
parametrization.  For example, 
%
the computation of the density of a $d$-dimensional Manhattan set
is enabled by a spatial partition whose $2^d$ atoms are indexed by $\bvec$'s.
Similarly, the  onion-peeling reconstruction
procedures mentioned previously are keyed to a partition of 
frequency space whose $2^d$ atoms are indexed by $\bvec$'s.
The frequency-domain version reconstructs the image
spectrum one atom at a time, beginning with ``highest frequency"
atoms (whose $\bvec$'s contain the most 1's), and working towards the lower frequency atoms
(whose $\bvec$'s contain fewer ones).

In particular, as will be shown, the spectrum $X^\bvec(\u)$ in the atom  indexed by $\bvec$ is computed via 
\begin{equation}  \label{eq:Xbreconstruct}
    X^{\bvec}(\u) ~=~ 
	X_\bvec(\u) \, - \!\! 
	\sum_{\bvec' : \,  \|\bvec^\prime\| > 	\|\bvec\|}  \hspace{-3.5ex}  X^{\bvec^\prime}_\bvec (\u)  \, ,
\end{equation}
where $X_\bvec(\u)$ is the spectrum of 
the image samples in the bi-step lattice 
parametrized by $\bvec$ (a subset of the Manhattan set),
the sum is over all $\bvec'$ with more ones than $\bvec$, and
$X^{\bvec'}_\bvec(\u) $ is the spectrum of the samples (taken with the same
bi-step lattice) of the image component $x^{\bvec'}(\t)$ 
corresponding to atom $\bvec'$,  which has previously been reconstructed.


A discrete-space version of this requires only DFTs of the subsampling of the 
Manhattan samples and the previously reconstructed image components specified
in the above, as well as summing and subtracting.  Then an inverse DFT
computes the newly reconstructed component.  Summing all  such
components yields the reconstructed image. 
 
The method characterized by \eqref{eq:Xbreconstruct}, and the discrete-space version thereof,
 can also be carried out in the spatial domain by applying the right-hand side of 
  \eqref{eq:Xbreconstruct} to the corresponding sampled images, rather than their spectra, 
  and then applying an ideal bandpass filter that extracts
 just the frequency component corresponding to atom $\bvec$.  The impulse responses of these filters will be given later.
As will be seen, these impulse responses depend on the $k_i$'s and $\lambda_i$'s,
but not the choice of bi-step lattices that comprise the Manhattan set.  Moreover,
the $\lambda_i$'s have only a simple spatial scaling effect on the filters.

%
%
Finally, we note that the development 
for three dimensions benefits greatly from the efficient parametrization
of bi-step lattices mentioned earlier, and that 
with such, it is possible to derive the M-sampling theorem
and reconstruction procedure
in arbitrary dimensions with essentially no additional effort or 
notation.


 


We conclude the introduction by relating the present work to previous work. 
Multidimensional sampling theorems, showing that images with certain
spectral support regions can be reconstructed from certain samplings sets,
appeared first for  
lattice sampling sets in Peterson and Middleton
\cite{peterson62sampling},
then later for unions of shifted  lattices, i.e., lattice cosets,  
\cite{gaarder72, 
marks86, cheung90image, cheung93multidimensional, faridani90, faridani94generalized, walnut96, 
venkataramani00perfect, venkataramani01optimal, behmard02sampling, 
behmard09efficient, unnikrishnan2012},
although they were not always described as such.  

The earliest work \cite{peterson62sampling, gaarder72} 
required the spectral support region and sampling set to be chosen so that the spectral replicas induced by sampling did not overlap, and consequently, 
reconstruction could proceed simply by lowpass filtering
the sampled image.  For example, the approach of \cite{gaarder72} could be used to
reconstruct images from M-samples.  However, it would require the images to be 
bandlimited to the Nyquist region of the \emph{coarse (rectangular) lattice}, which is
the intersection (rather than union) of the bi-step lattices comprising the Manhattan set.

Nonoverlapping spectral replicas were not required in later work 
\cite{marks86, cheung90image, cheung93multidimensional, faridani90, faridani94generalized, walnut96, 
venkataramani00perfect, venkataramani01optimal, behmard02sampling, 
behmard09efficient, unnikrishnan2012},
and more complex reconstruction procedures
were proposed.  
Though not specifically intended for images, a seminal contribution
stimulating a number of \edits advances in image sampling was the multichannel, generalized sampling introduced 
by Papoulis \cite{papoulis77generalized}.
For example, Papoulis' framework is broad enough to include all image sampling
schemes based on lattices and unions of shifted lattices.

 
One difference between the present work and much  past work is that we
focus on a particular sampling set, namely a Manhattan set, and seek a largest possible
frequency region such that any image bandlimited to such 
can be reconstructed from the samples.  In contrast, much of  the past work 
\cite{marks86, cheung90image, cheung93multidimensional, 
willis95optimal1, venkataramani00perfect, venkataramani01optimal} 
focused on a particular frequency support region and sought a smallest possible sampling set, 
constructed from lattices and shifts thereof, such that images bandlimited to this region 
could be reconstructed from such sampling sets. 
Nevertheless, some of the latter approaches could be used to 
reverse engineer reconstruction procedures and/or 
spectral support regions for Manhattan sets, as we now discuss.

One substantial line of past work applies to sampling sets that consist of a
sublattice of some specified \emph{base lattice}, 
together with some of its cosets, each of which is a shift of the sublattice by
some base lattice point.  
In this case, the subsampling corresponding to each coset (including the sublattice itself)
can be viewed  as a \emph{channel} in a Papoulis multichannel, generalized sampling scheme.
Consequently, the method of \cite{papoulis77generalized} can be applied.
This is the approach taken by Marks and Cheung 
\cite{marks86, cheung90image, cheung93multidimensional}.
Since a Manhattan set can be viewed as the union of what we earlier called the 
coarse (rectangular) lattice \edits and some number of its cosets with respect
to the \emph{dense (rectangular) lattice}, which contains all points $\tvec$
such that for each $i$, its $i$th coordinate is an integer multiple of $\lambda_i$,
the Papoulis-Marks-Cheung (PMC) approach can be applied to Manhattan sets.

In particular, Marks and Cheung   
%
%
focused on images with a given spectral support region
and an initial base sampling lattice  
such that the induced spectral replicas of this support region do not overlap.  
They then showed that cosets of some sublattice could be removed from the base  lattice
until the sampling density was minimal
(in the Landau sense) or approached minimal.  Their method involved (a)
partitioning the Nyquist region of the initial base lattice into atoms the size and shape of 
the Nyquist region of the sublattice, (b)
counting the number of atoms of this partition that are not overlapped by any spectral 
replica of the designated support region induced by the initial base sampling lattice, 
and (c) showing that this number of sublattice cosets
can be removed from the initial base lattice due to their samples being linearly
dependent on other samples.  If the atoms of the partition 
are too coarse 
to 
closely match the set of frequencies not contained in any spectral support replica,
then choosing a sparser sublattice will enable a finer partitioning, resulting in 
a higher fraction of the base
samples being removed, which allows the sampling rate to be reduced
until it equals or approaches the Landau minimum.  
 
With hindsight, one can apply their approach to a Manhattan sampling
set.  For simplicity, consider a 2D case and assume $k_1 = k_2 = k$.
Suppose images are bandlimited to the cross-shaped Manhattan region,
and let the initial base sampling lattice and the sublattice be the dense
and coarse rectangular lattices
mentioned earlier.  In this case, there are $k^2$ cosets of the sublattice
(including itself).  One can then see that in the partition of the Nyquist region
of the base/dense lattice into atoms having the size and shape of 
the Nyquist region  
of the coarse lattice, the number of atoms that are 
not contained in any sampled spectra is $k^2 - (2k-1)$.
Thus, it is possible to remove all but $2k-1$ cosets,
which is precisely the number of Manhattan samples 
in one $k \times k$ fundamental cell
of the coarse lattice.
Unfortunately, the PMC approach does not determine which cosets
can be removed, so it does not directly tell us if the Manhattan
samples are sufficient to recover an image.   
While it does provide a matrix invertibility test that one can apply
in any particular case to see if the Manhattan samples are sufficient, 
it is not clear
how to analytically establish that one can remove all but the Manhattan
samples in all cases.  It is also not clear how the PMC approach
would have lead to the discovery 
that the union of the Nyquist regions of 
the bi-step lattices is a reconstructable spectra support region for Manhattan sampling,
especially in dimensions three and above.
However, once it is known that the Manhattan samples are sufficient
for the spectral support region found in the present paper, then
the Papoulis approach will directly lead to a reconstruction algorithm.

As both the PMC and onion-peeling approaches involve partitioning  frequency space,
it is interesting to note that in dimension $d$ the PMC approach
requires a partition into $\prod_{i=1}^d k_i$ atoms, whereas 
the onion-peeling algorithm partitions into only $2^d$ atoms.
The smaller size of the latter partition 
is due to its being closely tailored to the specific
structure of Manhattan samples.

Similarly, in another line of work, Faridani \cite{faridani90} 
derived a sampling theorem and reconstruction formula 
for unions of shifts of one lattice. Given a spectral support region,
the reconstruction involves partitioning this region in a certain way 
and setting up and solving a sizable number of systems of linear equations,
assuming that the equations have a solution.
Since a Manhattan set can be viewed as the union of shifts of a lattice
(the coarse lattice) 
and since we know from the results of the present paper that it is possible
to reconstruct M-sampled images bandlimited to the Manhattan
region, one could presumably solve the resulting equations 
to obtain a reconstruction formula.    
While this is interesting, finding the partition and setting up the equations
can be difficult, especially in high dimensions.  Thus, as before, the onion-peeling approach proposed in this paper
is more natural, intuitive and straightforward to implement.

\iftoggle{comment}{\footnote{temp: Matt put this paragraph after Faridani as suggested.  
I think it is better now. We should reread and discuss. DN thinks it is good.}}{}
While the PMC and Faridani approaches could be used to derive a reconstruction
method for any Manhattan set, in their basic form, 
they do not provide direct closed form reconstruction methods,
as given for example in this paper.  That is, given sets of  $k_i$'s and  bi-step lattices, they  outline a procedure that could be followed in order to derive a reconstruction method.  
Then, when the $k_i$'s or  bi-step lattices are changed,
the procedure must be followed again, essentially from scratch\footnote{The method can be derived assuming unit $\lambda_i$'s and then spatially scaled 
for the actual $\lambda_i$'s.}. 
In contrast, the reconstruction methods given in the paper are closed form,
requiring just step-by-step following of the reconstruction formulas,
which depend explicitly on the $\lambda_i$'s, $k_i$'s and  bi-step lattices.
While it is conceivable that with enough work this alternative 
approach could be made closed form, it would appear to take
much additional work, especially to make it apply to arbitrary dimensions.

Behmard \cite{behmard09efficient} derived a sampling theorem and reconstruction formula
for unions of shifts of more than one lattice, which includes
M-sampling, as it is a more general setting than \cite{faridani90}.
However, the \emph{compatibility conditions} required to apply this
approach are not satisfied by M-sampling and the
Manhattan spectral support region.



Other work on sampling with unions of shifted lattices includes that of 
(a) Venkataramani and Bresler 
\cite{venkataramani00perfect, venkataramani01optimal}, 
which considered unions of shifted lattices in one dimension, and (b) Unnikrishnan and Vetterli \cite{unnikrishnan2012},  which 
considered unions of shifted lattices in higher dimensions.  The latter include M-sampling and a
reconstruction procedure was proposed with similarities to our onion-peeling approach,
but which requires the spectral support region to be convex, which rules
out the Manhattan region.  Indeed, one of their examples is a 2D
Manhattan grid, from which images can be recovered provided
their spectra are bandlimited to a circular subset of
the Manhattan region.  Consequently, a significantly smaller set of images
is reconstructable with their procedure.

Finally, we mention that Manhattan-bandlimited spectra have
been found to arise naturally in dynamic medical imaging applications,
including both time-varying tomography \cite{willis95optimal1}
and dynamic MRI \cite{rilling13multilattice}.  
For example, 
Rilling et. al. \cite[Fig. 1]{rilling13multilattice} give carotid blood velocity
mapping as an example of a dynamic MRI application where a cross-shaped
spectrum appears.
Moreover,
such spectra arise when temporal variation is
localized to a small spatial area relative to the rest of the body, such as
beating heart.  
With this motivation, Willis and Bresler 
\cite{willis95optimal1} derived a single sampling lattice such
that the cross-shaped spectral replicas did not overlap and the
sampling rate was close to the Landau lower bound.
In contrast, our sampling theorem also shows perfect reconstruction is possible.
However, we sample with more than one lattice, the spectral replicas overlap, 
and the Landau bound is met exactly.  \edits 
\iftoggle{comment}{\footnote{DN comment:  the statement about "our sampling theorem" was out of place.
This rewrite needs careful checking.}}{}
%

%


In summary, given that the present paper shows that 
images bandlimited to the union of the Nyquist regions of the 
bi-step lattices of a Manhattan sampling set can be perfectly reconstructed
from the Manhattan samples, there are probably a number of alternative
ways to derive reconstruction algorithms.  In the view of the authors,
the onion-peeling method, whose development was guided
by the specific structure of Manhattan samples, is
a natural and efficient reconstruction method
with a straightforward interpretation in frequency space.
It is also closed form in terms of the parameters of the Manhattan set.
In addition, the union-of-bi-step-lattice viewpoint taken in this paper
leads naturally to the hypothesis that the union of Nyquist regions 
is a support region of images that are reconstructable from Manhattan samples.
It is not known if other approaches would have lead investigators to this region. 
\iftoggle{comment}{\footnote{DN comment:  if the paper is overly long, this paragraph
could be merged with the concluding section}}{}

The paper is written so that the reader who is primarily interested in 2D images can focus on Sections \ref{sec:prelims}, \ref{sec:twoD}, and
\ref{sec:conclusions}.

\section{Preliminaries}\label{sec:prelims}

This section provides background and notation for sampling and lattices
that will be used throughout the the paper.

Let $\mathbb R$ denote the real numbers, let $\mathbb R^d$ denote
$d$-dimensional Euclidean space, let $\mathbb Z$ denote the set of all
integers, and let $\mathbb Z^d$ denote the set of all integer-valued
$d$-dimensional vectors.  In dimension $d$, an image is a mapping $x(\tvec) :
\mathbb R^d \to \mathbb R$, where the spatial variable is $\tvec = (t_1,
\ldots, t_d)$. 
We restrict attention to images $x(\t)$ that contain no delta functions or other generalized functions, and have well defined Fourier transforms 
containing no delta functions or other generalized functions,
%
%
where by Fourier transform we mean
$$ X(\u) ~=~ \mathcal F \big\{ x \big (\t) \big\} ~\triangleq~ \int x(\t) \,
e^{-j2\pi \t \cdot \u} \, d\t \, .  $$
We will often refer to $X(\u)$ as the \emph{spectrum} of $x(\tvec)$.

\emph{Sampling} a $d$-dimensional image $x(\tvec)$  means collecting its values
on some countable \emph{sampling set}  $\mathbb S$.  That is, it produces the
set of values $\{ x(\t) : \t \in \mathbb S \}$. 
 As commonly done, one can model such sampling 
 as multiplication of $x(\tvec)$ by the \emph{comb function} of the set $\mathbb
 S$, which produces the \emph{sampled image}
\begin{align}  \label{eq:xst}
   x_{\mathbb S}(\tvec) ~\triangleq~  x(\tvec) \, K_{\mathbb S} 
    \sum_{\tvec' \in  \mathbb S} \, \delta (\tvec -\tvec') \, ,   
\end{align} 
where $K_{\mathbb S}$ is a normalizing constant
and $\delta(\tvec)$ denotes the Dirac delta function in $d$-space.
The Fourier
transform of $x_{\mathbb S}(\tvec)$ is then called the \emph{sampled spectrum}.


{\em Rectangular sampling} refers to sampling with a rectangular lattice.
Given  $d$ and $\alphavec = (\alpha_1, \ldots, \alpha_d)$ with positive
components, the $d$-dimensional \emph{rectangular lattice} with \emph{step
vector} $\alphavec$ is a countably infinite set of points that are spaced by
integer multiples of the \emph{step size} $\alpha_i$ in the $i$th dimension.
Specifically, 
\begin{align*} 
L(\alphavec) 
   &~\triangleq~ \big \{ \tvec : t_i \mbox{ is a multiple of } \alpha_i, i=1,\ldots,d  \} \nonumber \\
   &~=~ \big \{ \tvec = \nvec \odot
  \alphavec:  \nvec \in \mathbb Z^d \} \, ,  
\end{align*} 
where $\odot$ denotes element-wise product (also known as the Hadamard or Schur product).
Alternatively, $L(\alphavec)$ is the additive group generated by 
the basis $\alpha_1 \evec_1, \ldots, \alpha_d \evec_d$, 
where $ \evec_1, \ldots,  \evec_d$ is the standard basis,
i.e., $\evec_i$ has a 1 in the $i$th place and 0's elsewhere.
%
%
%
%
%
%
That is,
\begin{equation*} L(\alphavec) ~\triangleq~ \Big \{ \tvec = \sum_{i=1}^d n_i
  \alpha_i \evec_i :  \nvec \in \mathbb Z^d  \Big\} \, . 
\end{equation*} 
%
The \emph{reciprocal lattice} corresponding to $L(\alphavec)$ is
\begin{equation*} L^*(\alphavec) ~\triangleq~ L(\alpha_1^{-1}, \ldots,
  \alpha_d^{-1}) \,.
\end{equation*} 
%

When sampling with set $\mathbb S =
L(\alphavec)$, it is convenient to set the normalizing constant in
(\ref{eq:xst}) to be 
\begin{align}   \label{eq:KS}
    K_{\mathbb S} \, = \,  {\textstyle \prod_{i=1}^d} \alpha_i  \, .
\end{align}
With this, 
the 
sampled image, denoted $x_{\alphavec}(\t)$,
has spectrum  
\begin{align}   \label{eq:sampledspectrum} 
  X_{\alphavec}(\u) 
 ~= \!  \sum_{\v  \in L^{^*}(\alphavec)}  \!\! X(\u - \v)   \, .
\end{align}
%
%
>From this, one sees that the sampled spectrum $X_{\alphavec}(\u)$ consists of
replicas of the original image spectrum $X(\u)$, translated to the sites in
frequency domain of the reciprocal lattice.  The usual $d$-dimensional sampling
theorem follows from the fact that if the support of \edits $X(\u)$ lies
entirely within the Nyquist region\footnote{In this paper, script
variables such as $\mathcal N, \B$ or $\A$ will usually denote subsets of frequency
space.} 
%
\begin{align*} \mathcal N_{\alphavec} ~\triangleq~ \Big\{ \u :  | u_i | < {1
  \over 2 \alpha_i}, \ i = 1, \ldots, d \Big \} \, , \end{align*} 
then said replicas do not overlap, and consequently, the original spectrum can
be recovered by extracting the portion of the sampled image spectrum in the
Nyquist region.

\section{Two-Dimensional Manhattan Sampling} \label{sec:twoD} 

As introduced earlier and depicted in Fig.\,\ref{fig:twodim}(a), Manhattan
sampling (M-sampling) uses locations spaced closely along a grid of horizontal and vertical
lines.  In particular, we assume there is a sample at the origin, as well as
samples spaced $\lambda_1$ apart on  horizontal lines spaced $k_2 \lambda_2$
apart, and samples spaced $\lambda_2$ apart on vertical lines spaced $k_1
\lambda_1$ apart, where $\lambda_i > 0$ and $k_1, k_2$ are integers greater
than one\footnote{We require $k_1,k_2> 1$ since if $k_1=1$ or $k_2=1$, the
sampling set reduces to a normal rectangular lattice. 
}.  The issue, now, is to
find an as large as possible set 
of images that can be perfectly
reconstructed from these samples, as well as an efficient procedure for doing
so.

A first thought is to model M-sampling as multiplying the given
image $x(\tvec)$ by a comb function having delta functions at the 
Manhattan sampling
locations, and then to analyze the spectra of the resulting sampled image.  Since
this comb function has the same periodicity as a comb function for
the \emph{coarse lattice} $L_C \triangleq L(k_1 \lambda_1, k_2 \lambda_2)$,
the replicas of the image spectrum
lie at frequency sites in the reciprocal lattice  $L^*_{C}$, or a subset
thereof.
  Thus, perfect reconstruction is possible for images bandlimited to
the Nyquist region ${\cal N}_{C}$ of the coarse lattice
$L_{C}$.  However, since such reconstructions need only use samples in the
coarse lattice, it may be that a larger set of images is
reconstructable from the full Manhattan grid.
On the other hand,  if images are bandlimited to a region larger than
$\mathcal N_C$, e.g., a scaling of the
Nyquist region such as $(1+\epsilon) \mathcal N_{C}$,  $\epsilon > 0$,
then the spectral replicas induced by an M-sampling comb may overlap.
Even if this does not eliminate the possibility of perfect reconstruction, it
will at least complicate the analysis.

Accordingly, we pursue an
%
%
approach that does not rely on nonoverlapping replicas, but derives from the key observation that the Manhattan
sampling set can be viewed as the \emph{union} of  two rectangular lattices.
Let us initially focus on what can be recovered from the samples 
of each lattice by itself.
The \emph{horizontal lattice},
$L_H \triangleq L(\lambda_1,k_2\lambda_2)$, densely samples in the horizontal direction and
coarsely samples in the vertical direction;  the \emph{vertical lattice},
$L_V \triangleq L(k_1\lambda_1,\lambda_2)$, coarsely samples in the horizontal direction and
densely samples in the vertical direction; and the sampling set for
M-sampling is 
\begin{equation*} M(\lambdavec;\kvec) ~=~ L_H  \cup
  L_V  \, .  
\end{equation*} 
Note also that the
  intersection of the two lattices is the coarse lattice $L_C $, whose
  comb function has the same periodicity as a comb function for the Manhattan
  grid.  

\begin{figure*}[!t] \centerline{ 
  \subfloat[]{\includegraphics[width=1.75in]{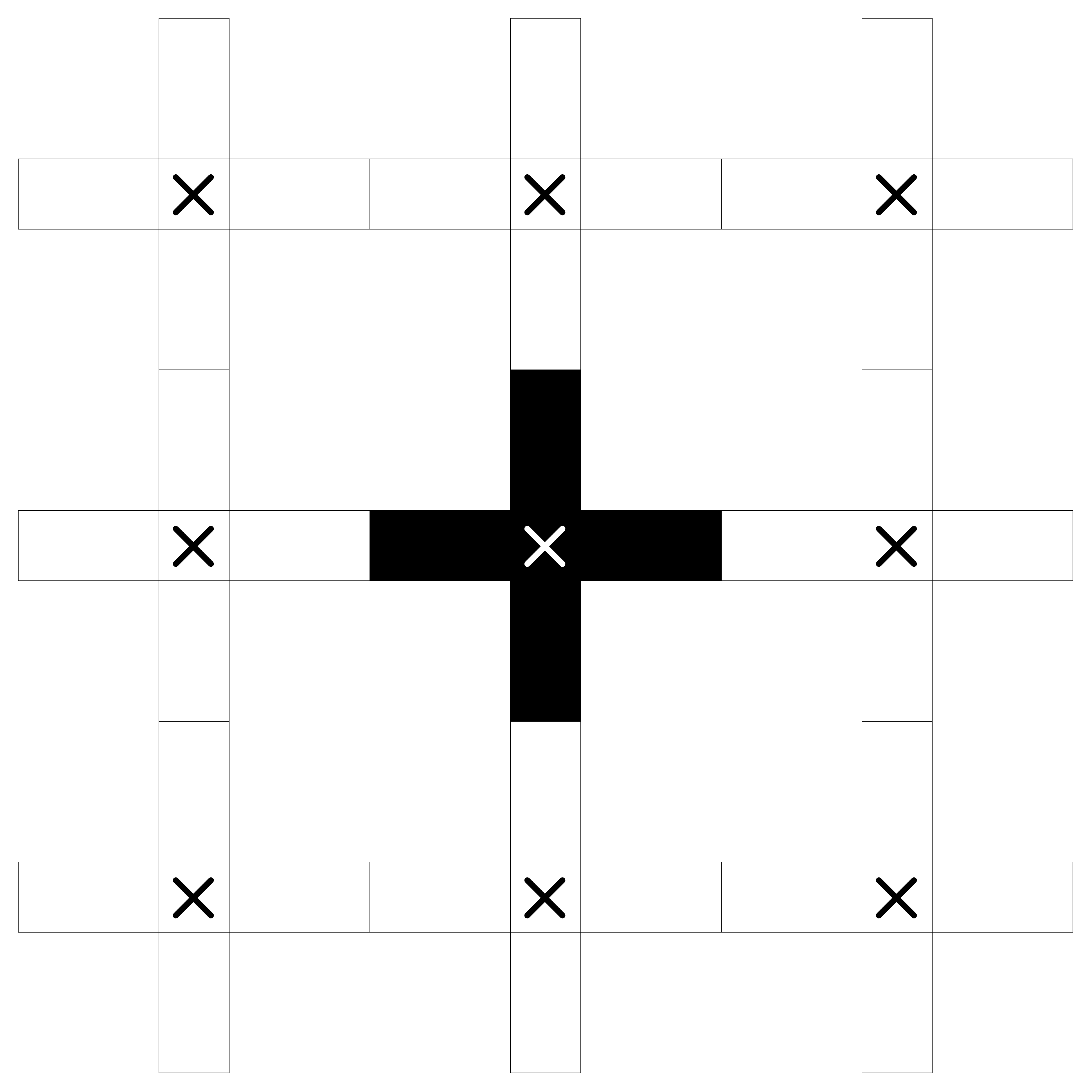}
  %
  }
  \hfil
  \subfloat[]{\includegraphics[width=1.75in]{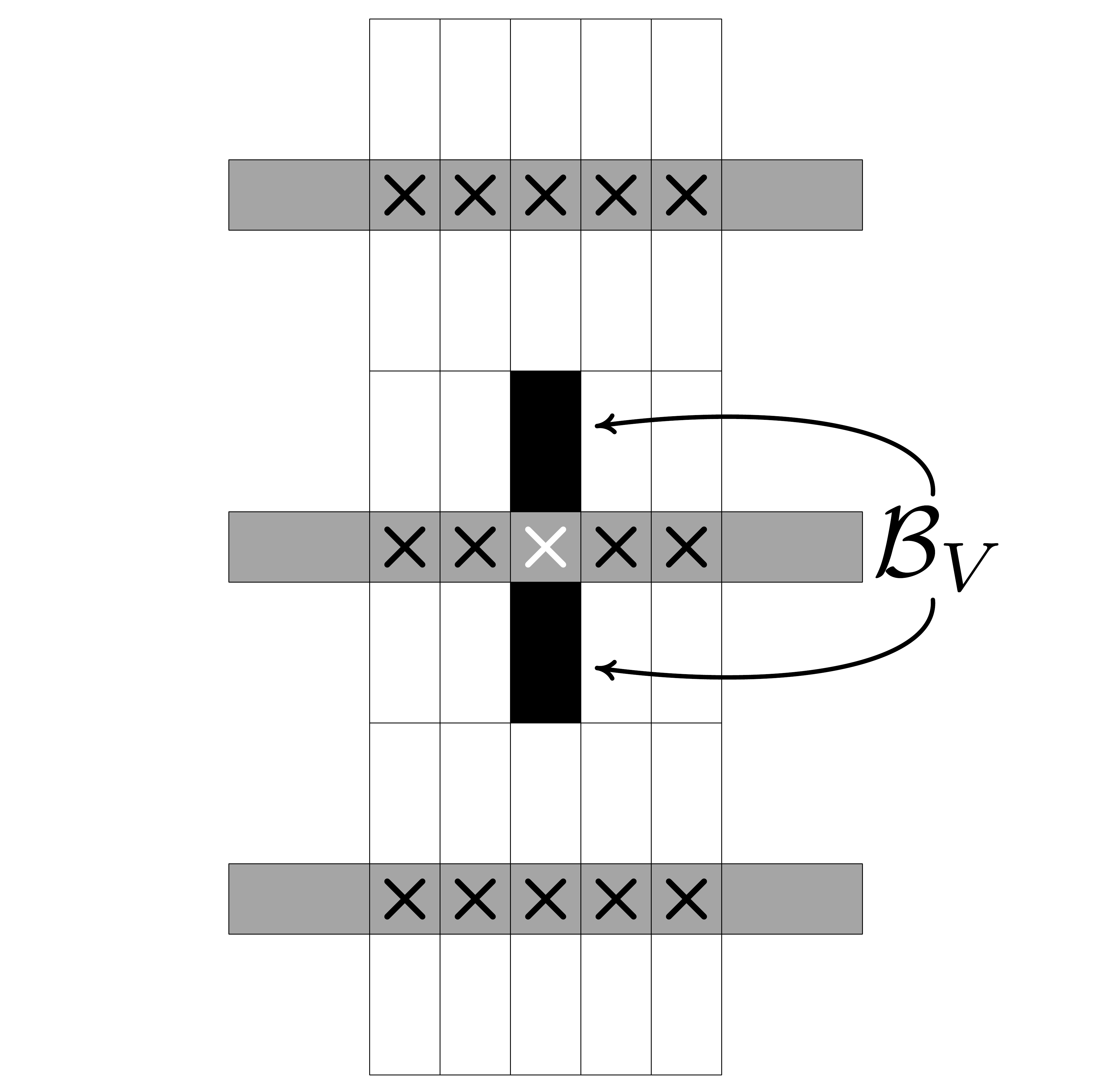}
  %
  }
  \hfil
  \subfloat[]{\includegraphics[width=1.75in]{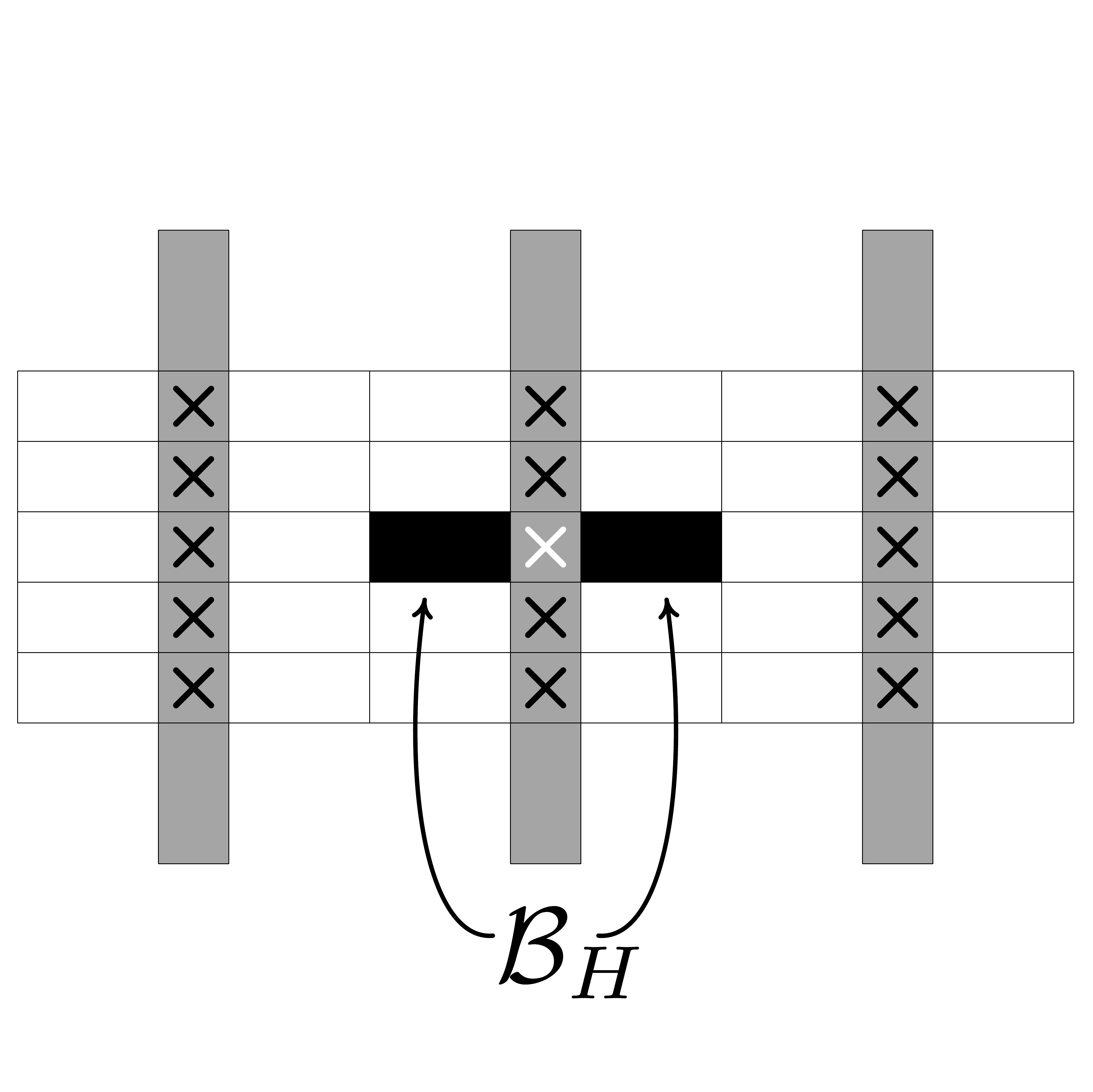}
  %
  }
  } 
  \caption{ For Manhattan sampling with $\lambda_1 = \lambda_2$ and $k_1=k_2
  =3$:  (a)  Support of the sampled spectrum for an image bandlimited to the
  Manhattan region, when sampled with the fine lattice
  $L_{\lambda_1,\lambda_2}$.  The original spectrum is black with
  a white $\times$ in its center, whereas replicas are white with a black
  $\times$ in their centers.   
 (b) Support of the sampled spectrum  when
  sampled with the vertical lattice $L_{k_1 \lambda_1,\lambda_2}$.
  Gray indicates regions where replicas overlap either the original
  spectrum or each other. 
  %
  (c) Same as (b), except that the sampling is with the
  horizontal  lattice $L_{\lambda_1,k_2 \lambda_2}$.
  %
  }
  \label{fig:alias} 
\end{figure*}

Clearly, all images bandlimited to the
Nyquist region ${\cal N}_H$ of the horizontal
lattice  $L_H$ 
can be recovered from just the samples
in this lattice.  
Likewise, all images bandlimited to the
Nyquist region ${\cal N}_V$  of the vertical 
lattice $L_V$
can be recovered from just the samples
in this lattice. 
Each of these by itself leads to a larger recoverable set of images than
the set recoverable from sampling with the coarse lattice $L_C$.  
However, neither type of sampling and reconstruction uses all of the M-samples.


We now show how  images bandlimited to the union of 
the Nyquist regions of the horizontal and vertical 
lattices can be recovered from the full set of M-samples.
Specifically, suppose image  $x(\t)$ is bandlimited to  ${\cal M}(\lambdavec;\kvec)=
{\cal N}_H \cup {\cal N}_V$,
 which is the cross-shaped region shown in Fig.\,\ref{fig:twodim}(c).  
First, consider only the samples of $x(\t)$ taken on the vertical 
lattice. Since the cross-shaped region ${\cal M}(\lambdavec;\kvec)$ is not
contained in the Nyquist region ${\cal N}_V$,
the replicas of $X(\u)$ may overlap in the spectrum of the vertically 
sampled image,  as illustrated in Fig.\,\ref{fig:alias}(b),
However, certain portions of each cross-shaped
replica cannot be overlapped, and thus these portions of the spectra
of $x(\t)$ can be immediately recovered.

Specifically, it is easy to see that with  vertical sampling,
the vertical highpass region  $\B_V \triangleq  \Ncal_V - \Ncal_C$
is not overlapped.
Thus, with $I_D(\u)$ denoting the
indicator function of some set $D$ and $X^V(\u) \triangleq X(\u) I_{\B_V}(\u)$ denoting
the portion of $X(\u)$ in $\B_V$, one sees that 
from the vertical samples and their spectrum $X_V(\u)$, 
one can immediately recover $X^V(\u)$ 
via $X^V(\u) = X_V(\u) I_{\B_V}(\u)$.
%
Likewise from the horizontal samples and their spectrum $X_H(\u)$, the horizontal highpass region  $\B_H \triangleq  \Ncal_H - \Ncal_C$ 
is not overlapped.
Thus, one can immediately recover 
$X^H(\u) \triangleq X(\u) I_{\B_H}(\u) = X_H(\u) I_{\B_H}(\u)$.\footnote{Throughout the paper,
a superscript on an image $x$ or spectrum $X$ will usually pertain to
a frequency region, and a subscript will usually pertain to a sampling.}

%


Since $X^V(\u)$ and $X^H(\u)$ are now known, 
and $X(\u) $ is bandlimited to $\mathcal M(\lambdavec,\k) =
\B_H \cup \B_V \cup \mathcal N_C$,  
it remains only to find $X^C(\u)  \triangleq X(\u) I_{\mathcal N_C}(\u)$.
It will then follow that  
$ X(\u) = X^V(\u) + X^H(\u) + X^C(\u)$. 
Inverse transforms
will give 
$ x(\t) = x^V(\t) + x^H(\t) + x^C(\t)$.

  
To determine 
$X^C(\u)$, 
consider the vertical sampling of $x(\t)$, and observe in 
Fig.\,\ref{fig:alias}(b) that the overlap of the image spectrum $X(\u)$ in
${\cal N}_C$ by the various spectral replicas
is due only to
replications of the horizontal highpass frequency components 
in $\B_H$.  Since these have already been determined, it ought to be possible
 subtract their effects.  
 

 To see that this can be done, 
let us focus on $X_V(\u) \, I_{\mathcal N_C}(\u)$.
>From (\ref{eq:sampledspectrum}) and the fact that
$X(\u) =0$ for $\u \not \in \mathcal M(\lambdavec,\k)$, we have 
\begin{align*} 
  X_V(\u) \, I_{\mathcal N_C}(\u) ~ =  \sum_{i=-n}^n X \Big( u_1 - {i \over k_1\lambda_1}, u_2 \Big) \, I_{\mathcal N_C}(\u)  \, ,
\end{align*}
where $n = \big \lfloor {k_1 \over 2} \big \rfloor$.  
Now using  $X(\u) = X^V(\u) + X^H(\u) + X^C(\u)$ in the above 
along with 
 the facts that\\[1ex]
\indent (a) $X^V \big( u_1 - {i \over k_1\lambda_1}, u_2 \big) 
$ $	I_{\Ncal_C} (\u) = 0$ for all  $i$,  \\[1ex]
\indent (b) $X^H \big( u_1 - {i \over k_1\lambda_1}, u_2 \big) 
\, I_{\Ncal_C} (\u) = 0$ for   $i=0$, \\[1ex]
 \indent  (c) $X^C \big( u_1 - {i \over k_1\lambda_1}, u_2 \big) 
\, I_{\Ncal_C} (\u) = 0$ unless $i=0$, \\[1ex]
%
we find 
\begin{align}  \label{eq:SV2}
  X_V  (\u)  \, I_{\Ncal_C}(\u)  
         ~=~  X^C(\u) \,  + \,  Y(\u) \, I_{\Ncal_C}(\u) \, ,
\end{align} 
where
\begin{align}  \label{eq:Y}
        Y(\u)  &\,\triangleq    \sum^n _{\parbox{.3in}{ $\scriptstyle i=-n$\\[-2ex]
         $\scriptstyle i \neq 0$}} 
       \!\!\!  
        X^H \! \Big( u_1 - {i \over k_1\lambda_1}, u_2 \Big)  \notag \\
        &\,=   \sum^n _{\parbox{.3in}{ $\scriptstyle i=-n$\\[-2ex]
         $\scriptstyle i \neq 0$}} 
       \!\!\!  
        X_H \! \Big( u_1 - {i \over k_1\lambda_1}, u_2 \Big)
        I_{\B_H} \! \Big( u_1 - {i \over k_1\lambda_1}, u_2 \Big)     .
\end{align} 
where the last equality uses the fact, mentioned earlier, that $X^H(\u)= X_H (\u) I_{\B_H}(\u)$.
Notice that $Y(\u)$ is the component of $X_V(\u)$ due to aliasing by
replicas of $X^H(\u)$, and is directly computable from the 
horizontal samples. 
%
It follows from \eqref{eq:SV2} that $X^C (\u) = ( X_V(\u) - Y(\u) ) \, I_{\mathcal N_C}(\u) $.

In summary, a procedure for recovering a cross bandlimited $x$ from
its M-samples is
\begin{enumerate}
\item  Compute the spectra, $X_H(\u)$ and $X_V(\u)$, of the horizontally and vertically dense samples, respectively.

\item From $X_H(\u)$, compute $Y(\u)$ for $\u \in \mathcal N_C$.  

\item   Let  
\begin{align}  \label{eq:Xhatu}
     \widehat X (\u) = \begin{cases}
                    X_V(\u),  &  \u \in \B_V \\[0ex]
                   X_H(\u), & \u \in \B_H    \\[0ex]
                   X_V(\u) - Y(\u) , & \u \in {\cal N}_C
                      \end{cases}  .
\end{align}

\item  Let  $\widehat x(\t)$ be the inverse Fourier transform of $\widehat X(\u)$.
\end{enumerate}

This result is summarized in the following.

\vspace{1ex}
\begin{theorem} \emph{2D Manhattan sampling theorem. }
\label{thm:2Dreconstruction}
Given $\lambda_1, \lambda_2 >0$ and integers $k_1, k_2$ greater than
1, any  image $x(\t) $ whose Fourier transform is bandlimited to the cross-shaped region ${\cal M(\lambdavec; \kvec)}$
can be recovered from its M-samples in 
$M(\lambdavec;\kvec)$ with the procedure given above.
\end{theorem}

\vspace{1ex}

The following alternative expression for $X_V(\u) \, I_{N_C}(\u)$
will lead to an easier to implement procedure for discrete-space images 
with finite support (presented later).  
Using  (\ref{eq:sampledspectrum}) and $X(\u) = X^V(\u) + X^H(\u) + X^C(\u)$, we find
\begin{align*}
    X_V(\u) \, I_{\mathcal N_C} (\u) 
      &=   \sum_{\v \in L^*_V} \!\!\! \Big( X^V(\u - \v) + X^H(\u - \v) + X^C(\u - \v) \Big) I_{\mathcal N_C} (\u) \\
      &= \, X^C (\u) \, +  \sum_{\v \in L^*_V} \!\!  X^H(\u - \v) \ I_{\mathcal N_C} (\u)  \\
      & =  \,  X^C (\u) \, +  \, Y'(\u) I_{\mathcal N_C}(\u)  \, ,
\end{align*}
where 
\begin{align}  \label{eq:Y'}
    Y'(\u) ~\triangleq \sum_{\v \in L^*_V} \!\!  X^H(\u - \v)  \, .
\end{align}
It follows that $Y(\u)$ in the procedure given previously can be replaced by $Y'(\u)$.
The advantage is that, as shown below, $Y'(\u)$ can be computed 
with Fourier transforms instead of a summation. 
To show this, let $\mathcal S_V$  denote the vertical sampling
operator, which when applied to an image $z(\t)$ produces 
$z_{L_V}(t)$ 
as defined by  (\ref{eq:xst}).  We
recognize the summation in (\ref{eq:Y'}) as the sampled spectrum when the
image $x^H(\t)$ is vertically sampled. Since $x^H(\t)$, and consequently
$X^H(\u)$, can be computed from the horizontal samples,  
\begin{align*}  \label{eq:Y'}
      Y'(\u) 
           &~=~    \mathcal F \big\{ \mathcal S_V \big\{ x^H(\t) \big\} \big\}   
              ~=~   \mathcal F \big\{ \mathcal S_V \big\{ \mathcal F^{-1} \big\{ X^H(\u) \big\} \big\} \big\}     \\[.5ex]
           &~=~   \mathcal F \Big\{ \mathcal S_V \Big\{ 
                    \mathcal F^{-1} \big\{  I_{\mathcal B_H} (\u)
                             \mathcal F \big\{   
                                x_H(\t)             
                             \big\}                 
                   \big\} \Big\}
         \Big\}  \, .  
\end{align*}
While the above may initially appear complex\footnote{Note also that
the expression (\ref{eq:Y'}) for $Y'(\u)$ contains more terms in the sum than
the corresponding expression (\ref{eq:Y}) for $Y(\u)$.}, in the discrete-space, finite-support case
discussed shortly, it will lead to a simple procedure that avoids
the summations in (\ref{eq:Y}) and (\ref{eq:Y'}).

\subsection*{Maximality, in the Landau sense,  of the set of reconstructable images} 

 
The sampling density of an
$M(\lambdavec;\kvec)$ M-sampling set is 
\begin{equation*}
  \rho = \frac{k_1+k_2-1}{k_1 k_2 \lambda_1 \lambda_2} \, ,
\end{equation*}
since any $k_1 \lambda_1 \times k_2 \lambda_2$ rectangle in $\mathbb R^2$
contains $k_1+k_2-1$ samples.  In the frequency domain, the area of
the Manhattan-bandlimited region, denoted $| \M (\lambdavec; \kvec) |$,  is the sum of the areas
of $\B_H$, $\B_V$ and ${\cal N}_C$.  Alternatively, it is sum of the areas of the Nyquist regions corresponding to the horizontal and vertical
sampling lattices, minus the area of their intersection.  Either way, this may be written
as
\begin{equation*}
  |\B_H|+|\B_V|+|{\cal N}_C| = 
  \frac{1}{k_1 \lambda_1 \lambda_2} + \frac{1}{k_2 \lambda_1 \lambda_2} -
  \frac{1}{k_1k_2\lambda_1\lambda_2} \, ,
\end{equation*}
which simplifies to  the previous expression for sampling density $\rho$. \edits  Thus, the
set of images bandlimited to the Manhattan region $\mathcal M(\lambdavec; \k)$
is a maximal set of reconstructable images  in the Landau sense for the M-sampling grid $M(\lambdavec; \k)$.

\subsection*{Discrete-space images} \label{sec:discretespace}

In this section, we briefly consider M-sampling of discrete-space
images. Such images might be created by rectangularly sampling a
continuous-space image, or they might exist only as discrete-space objects.  In
any case, we consider an image to be a mapping $x[\t] : \mathbb T \to \mathbb
R$ where $\mathbb T$ is either the (infinite) integer lattice $\mathbb Z^2 $, 
or a finite subset of the form $\mathbb T = \{ \t : 0 \leq t_1 \leq T_1 -1 , \, 0 \leq t_2 \leq T_2 -1 \}$
for some positive integers  $T_1, T_2$.  
%
\emph{Sampling} $x[\t]$ refers to collecting a subset of its values.  
In the infinite support case, $\mathbb T = \mathbb Z^2$, \edits a
Manhattan grid $M(\lambdavec, \k)$ is once again defined to be the 
union of a horizontal lattice
$L_H \triangleq L(\lambda_1, k_2 \lambda_2)$ and a vertical lattice
$L_V \triangleq L(k_1 \lambda_1, \lambda_2)$, 
except that now each lattice must be a sublattice of the integer lattice $\mathbb
Z^d$, i.e., $\lambda_1,\lambda_2$ must be positive integers.  
In this case, we assume the discrete-space Fourier transform of $x[\t]$ is well defined 
and contains no delta functions or other generalized functions.
In the finite support case, a Manhattan grid is formed in a similar way, namely, 
$M(\lambdavec, \k) = L_H \cup L_V$, where now $L_H$ 
and $L_V$ are truncated to the finite $\mathbb T$.

\vspace{1ex}
\emph{(a) Infinite-support discrete-space images:}
In this case, Theorem \ref{thm:2Dreconstruction} holds
with only trivial changes, as does the 
reconstruction procedure.  Specifically, the only required changes are:
(i) replace the continuous-space Fourier transform as the formula for
a spectrum with the discrete-space Fourier
transform, and (ii) scale  all specified frequencies by $2 \pi$, such as those defining Nyquist regions and $\mathcal M(\lambdavec; \k)$.

\vspace{1ex}

%

\emph{(b) Finite-support discrete-space images:}
In this case, as is customary, we use the Discrete Fourier Transform (DFT) as
the formula for the spectrum of an image: 
\[ 
    X[\u] ~= \sum_{\t \in \mathbb T}
      x[\t] \, e^{-j 2\pi ({u_1 \over T_1}t_1 + {u_2 \over T_2} t_2 )},  ~~ \u \in
       \mathbb T \, .  
\] 
%

The conventional sampling theorem (c.f.\,\cite{behmard09efficient})  
for discrete-space images 
with spatial support  $\mathbb T$ (defined by $T_1, T_2$)  sampled with a
rectangular lattice $L(\alpha_1, \alpha_2)$ limited to $\mathbb T$
 says that
an image $x[\t]$ with support $\mathbb T$  can be recovered from its samples
in this lattice if $T_1$ and $T_2$ are integer multiples of $\alpha_1$ and 
$\alpha_2$, respectively, and its DFT $X[\u]$ is zero outside the (discrete) Nyquist region 
\begin{align*}  
      \widetilde{ \mathcal N}_{\alpha_1,\alpha_2} ~\triangleq~& \Big\{ \u \in \mathbb T:
             \mbox{ for }i = 1 \,  \& \, 2, ~ 0 \leq u_i < {T_i \over 2 \alpha_i}  
            \mbox{ or } T_i - {T_i \over 2 \alpha_i} < u_i \leq T_i-1 \Big \} \, .
\end{align*}

Now suppose a finite-support discrete-space image $x[\t]$ is sampled 
on the Manhattan grid $M(\lambdavec, \k)$
and is bandlimited to the cross-shaped \emph{Manhattan region}
$$
      \widetilde{\mathcal M}(\lambdavec, \k) ~\triangleq~ 
         \widetilde{\mathcal N}_{H} \, \cup \,  \widetilde{\mathcal N}_{V} \, ,
$$
where $\widetilde{\mathcal N}_{H}$ and $\widetilde{\mathcal N}_{V}$ are
the Nyquist regions of the horizontal and vertical lattices, respectively.
Assuming $T_1$ and $T_2$ are integer multiples of $k_1 \lambda_1$ and
 $k_2 \lambda_2$, respectively, a straightforward adaptation of the analysis for 
continuous-space images shows that 
from the samples in the vertical lattice
$L_V$, 
one can recover the spectrum $X[\u]$ in the highpass region 
$\Btilde_V \triangleq \Ntilde_V - \Ntilde_C$,
where $\Ntilde_V$ and $\Ntilde_C$ are the Nyquist regions of the 
vertical and coarse lattices, respectively.
Specifically, from the spectrum $X_V[\u]$ of the vertically sampled
image $x_V[\t]$ (with scaling as in (\ref{eq:xst})-(\ref{eq:KS})),  one recovers
$X^V[\u] \triangleq X[\u] I_{\Btilde_V}[\u] = X_V[\u] I_{\Btilde_V}[\u]$.
Likewise, from the  samples in the horizontal lattice
 $L_H $, 
one can recover the spectrum in the highpass region 
$\Btilde_H \triangleq \Ntilde_{H} - \Ntilde_C$ 
from the spectrum $X_H[\u]$ of the horizontally sampled image $x_H [\t]$  via
$X^H[\u] \triangleq X[\u] I_{\Btilde_H}(\u) = X_H[\u] I_{\Btilde_H}[\u]$.
Finally, the spectrum in the Nyquist region $\Ntilde_C$ of the coarse
lattice can be determined via  $X^C [\u] = ( X_V[\u] - Y[\u] ) \, I_{\Ntilde_C}[\u] $,
where
\[
       Y[\u] ~=~ k_1 \lambda_1\lambda_2  \sum_{r=1}^{ k_1 \lambda_1 -1}
   X^H  \Big[\Big(u_1 - r {T_1 \over k_1 \lambda_1} \Big) \bmod T_1, u_2  \Big] \, .
\]
This leads to the following.

\vspace{1ex}
\begin{theorem}\emph{2D discrete-space, finite-support Manhattan sampling
  theorem. }     \label{thm:finite_discrete}   
 If  $T_1$ and $T_2$ are integer multiples of $k_1 \lambda_1$ and
 $k_2 \lambda_2$, respectively, then an image $x[\t]$ with finite support $\mathbb T$ 
 can be recovered from its M-samples in $M(\lambdavec,\k)$ 
 if its DFT $X[\u]$ is zero outside the Manhattan region
 $\Mtilde(\lambdavec, \k)$.  
\end{theorem}
\vspace{1ex}

\emph{Reconstruction procedure:} 
%

Given the samples in Manhattan grid $M(\lambdavec,\k)$ of an image $x[\t]$
bandlimited to $\widetilde{\mathcal M} (\lambdavec,\k)$, the following
adaptation of the continuous-space procedure recovers the entire $x[\t]$.  
%

\begin{enumerate}

\itemsep=3pt

\item Let $x_V[\t]$ equal  $k_1\lambda_1 \lambda_2 \, x[\t]$ on the vertical lattice
 $L_V$ and zero otherwise, and 
 let  $x_H[\t]$ equal  $k_2 \lambda_1\lambda_2 \, x[\t]$ \edits
 on the horizontal lattice
 $L_H$ and zero otherwise. 
 Compute $X_V[\u] = \DFT \big \{x_V[\t] \big\}$ and  $X_H[\u] = \DFT \big \{x_H[\t] \big\}$.

\item Compute the ``alias subtraction'' term
\[
   Y' [\u] ~=~    \DFT \Big\{ \widetilde{\mathcal S}_V \Big\{ 
                                \IDFT \big\{  
                              I_{\Btilde_H} [\u] \,  \DFT \{     x_H [\t]   \}             
                   \big\} \Big\}
         \Big\}  ,
\]                       
where   $\widetilde{\mathcal S}_V$ denotes the vertical sampling operator that,
when applied to an image $z[\t]$, 
produces 
an image that is 
$ k_1\lambda_1 \lambda_2 \, z[\t] $ on 
$L_V$, and zero elsewhere.                            
                               
%
   %
%
%

 \item Compute the spectrum:   
  \[
      \widehat X[\u] ~=~ 
      \begin{cases}
         X_V[\u], & \u \in \widetilde \B_V\\
        X_H[\u], & \u \in \widetilde \B_H\\
        X_V[\u] - Y'[\u], & \u \in \widetilde{N}_C 
   \end{cases}  \, .
\] 
 \item Invert the spectrum:
\[     
    x[\t] ~=~ \IDFT \big\{X[\u] \big\}  \, .
 \]
\end{enumerate}

This reconstruction procedure uses 5 DFT/IDFT operations, each
requiring $\mathcal O(N \log N)$ arithmetic operations when implemented with an
FFT,  where $N = T_1 T_2 $, plus three pairwise additions of $T_1 \times T_2$
matrices, each requiring $T_1 T_2$ additions, plus instances of setting matrix
elements to zero.  In summary, the complexity of reconstruction, which is
dominated by the FFT's, is $\mathcal \O(N \log N)$ operations per
image. 

%
\begin{figure}[!t] 
  \centering 
  \subfloat[]{\includegraphics[width=1.2in]{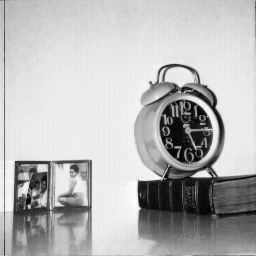} } %
  \subfloat[]{\includegraphics[width=1.2in]{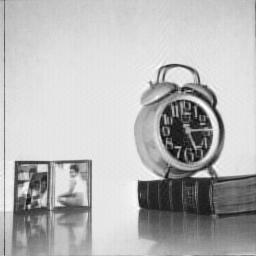} }
  \subfloat[]{\includegraphics[width=1.2in]{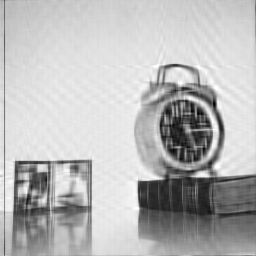} }
  \subfloat[]{\includegraphics[width=1.2in]{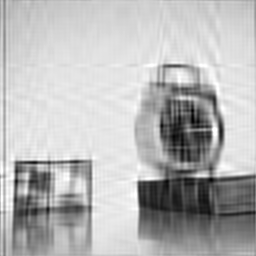} }
  \subfloat[]{\includegraphics[width=1.2in]{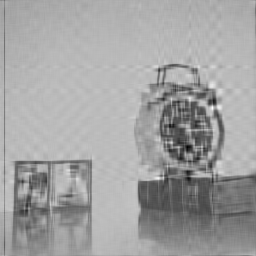} }
  \\
  \subfloat[]{\includegraphics[width=.83in]{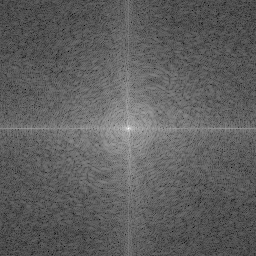} }
  \subfloat[]{\includegraphics[width=.83in]{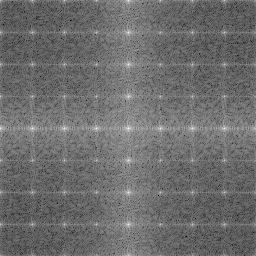} }
  \subfloat[]{\includegraphics[width=.83in]{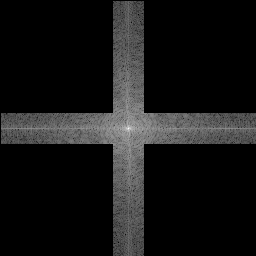} }
  \subfloat[]{\includegraphics[width=.83in]{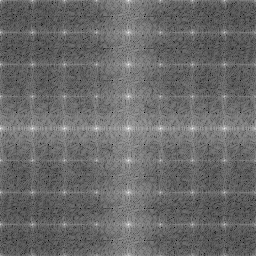} }
  \caption{(a) Original $256 \times 256$ image. (b) Image 
  bandlimited to
  Manhattan region $\widetilde{\mathcal
  M}(\lambdavec,\k)$, with $k_1 = k_2=4$ and $\lambda_1=\lambda_2=1$.
  (c) Same as (b) except $k_1=k_2=8$.  
  (d) Same as (c) except $\lambda_1=\lambda_2=2$.  
(Note:  after spectra were zeroed outside
  $\widetilde{\mathcal M}(\lambdavec,\k)$, inverse transforms were applied,
  negligible imaginary parts were discarded, and images were quantized
  to $\{0, 1, \ldots, 255\}$.)  
  (e) Image sampled with parameters of (c) and reconstructed without first bandlimiting to Manhattan region.
  Log magnitude spectra: (f) original image; (g) original image sampled with parameters
  of (c); (h) original image bandlimited with parameters of (c); 
  (i) bandlimited image (c) sampled with parameters of (c);
}
  \label{fig:image_examples} 
\end{figure}
%

Note that few real world, finite-support images will satisfy the conditions of Theorem
\ref{thm:finite_discrete}.  As a result,  to apply M-sampling 
to a real world image, the image can be pre-processed by
zero-padding so that its dimensions are multiples of $k_1 \lambda_1$ and
$k_2 \lambda_2$, respectively, and ``Manhattan filtering'' \edits by taking
the DFT and setting to zero all coefficients 
 outside of  $\widetilde{\mathcal M} (\lambdavec, \k )$.  
 Such padded and filtered images 
 can be recovered perfectly
from their M-samples.  To illustrate the effects of such filtering, 
which heavily suppresses diagonal frequencies, Fig.\,\ref{fig:image_examples} shows a finite-support image and its filtering with several choices of
parameters.  It also shows the spectra of the image before and after
bandlimiting, the spectrum of the sampled image, with and without
bandlimiting, and the effect of sampling and reconstruction without first pre-filtering.  Note that  
the image was chosen to have sharp edges
surrounded by a smooth background in order that one can easily
see the ringing due to bandlimiting. 
\iftoggle{comment}{{\textbf{\textit{aren't we going to add an image showing
the result of reconstructing without pre-filtering in one of the cases?}}}}{}

%

%
\section{Higher-Dimensional Manhattan Sampling}
\label{sec:higherD}

\subsection{Introduction}

As mentioned earlier, in any dimension $d \geq 3 $ there are a number of
possible $d$-dimensional Manhattan sets.  
Each is
a finite union of rectangular lattices, each defined by step sizes that in
dimension $i$ are constrained to be $\lambda_i$ or $k_i \lambda_i$, where each
$\lambda_i$ is a positive constant called the \emph{dense spacing} in dimension
$i$, and each $k_i$ is an integer greater than 1 called the \emph{sampling
factor} in dimension $i$.  Such a rectangular lattice will be called a
$(\lambdavec, \kvec)$-lattice, where $\lambdavec = (\lambda_1, \ldots,
\lambda_d) $ is its \emph{dense spacing vector} and  $\kvec = (k_1, \ldots,
k_d)$ is its \emph{sampling factor vector}.  It will also be called a $(d,
\lambdavec, \kvec)$-lattice when we wish to emphasize $d$, and a \emph{bi-step
lattice} when we do not wish to specify parameters.  (The term ``bi-step'' 
emphasizes that each step size $\alpha_i$ can only take one of two values:
$\lambda_i$ or $k_i \lambda_i$).
%
%
Accordingly, to specify a $d$-dimensional \emph{Manhattan set}, one specifies a dense
spacing vector  $\lambdavec $, a sampling factor vector $\kvec$, and a
collection of $(\lambdavec, \kvec)$-lattices. 



To efficiently characterize a $(\lambdavec, \kvec)$-lattice, we let $\bvec = (b_1, \ldots, b_d)$
denote a $d$-dimensional vector, called its \emph{bi-step indicator vector}, or more concisely 
\emph{bi-step vector},
that indicates the dimensions along
which the bi-step lattice is dense, according to the convention $b_i = 1$ if the step size is
$\lambda_i$  in dimension $i$ and
0 if  the step size is $k_i \lambda_i$.
Thus, the $(\lambdavec, \kvec)$-lattice specified by $\bvec$ is 
$L_{\lambdavec, \kvec, \bvec} \triangleq L(\alphavec_{\bvec})$, with $\alphavec_{\bvec} =
  (\alpha_{\bvec,1}, \ldots, \alpha_{\bvec,d})$ defined by 
 \begin{align}
    \label{eq:alphagamma} \alpha_{\bvec,i} 
         ~=~ \begin{cases} \lambda_i, & b_i = 1\\ k_i \lambda_i, & b_i=0 
         \end{cases} \, .  
\end{align}  
or equivalently, 
\begin{align}
  L_{\lambdavec, \kvec, \bvec} 
   & \,\triangleq\, \big\{  \t :  t_i \mbox{ is a multiple of $k_i \lambda_i$  for $i$ s.t. $b_i=0$,}
           \mbox{ and a multiple of } \lambda_i   \mbox{ for other } i \big\}  \, .  \label{eq:Llambdakb}  
 \end{align} 
Note that we generally consider $d$, $\lambdavec$ and $\kvec$ to be fixed, and
so as an abbreviation and slight abuse of notation, we usually write
$L_{\bvec}$ instead of $L_{ \lambdavec, \kvec, \bvec}$.  It will also be useful to let  $x_\bvec(\t)$ and $X_\bvec(\u)$
denote, respectively, the sampled image and the sampled spectrum
due to sampling $x(\t)$ with $L_\bvec$.

%
The following summarizes.

\vspace{1ex}
\begin{definition} 
Given dimension $d$, dense spacing vector $\lambdavec$, 
sampling factor vector $\kvec$ (all of its components are integers greater than 1), and a finite collection of
$(d, \lambdavec, \kvec)$-lattices
specified by the bi-step vectors in $B = \{ \bvec_1, \ldots \bvec_m \}$, 
the corresponding $(d, \lambdavec,\kvec, B)$-\emph{Manhattan (sampling) set} is
\begin{align}  \label{eq:M-grid}
      M \big(d, \lambdavec, \kvec, B \big) ~\triangleq~  
          \bigcup_{j=1}^m L_{\bvec_j} \, .
\end{align}
\end{definition}
As $d, \lambdavec$ and $\kvec$ will be considered fixed,
we usually write $M\big(B\big)$ instead of $ M\big(d, \lambdavec, \kvec, B\big)$.  $B$ will be called a \emph{Manhattan 
collection} or \emph{M-collection} for short.
%
%

\subsection{Examples and properties of bi-step lattices} 

It is useful to call attention to certain bi-step lattices.
%
%
One is the \emph{dense lattice} $L_{\mathbf 1}$
corresponding to the bi-step vector $\bvec = \mathbf 1 \triangleq (1, \ldots, 1)$.  
It is a rectangular lattice with step
vector $\lambdavec$ that contains every other $(\lambdavec,
\kvec)$-lattice.
Another is the \emph{coarse lattice} $L_{\mathbf 0}$ corresponding to $\bvec =
\mathbf 0 \triangleq (0, \ldots, 0)$, which is the rectangular   lattice with step vector   
$\alphavec = \kvec \odot
\lambdavec$ and  which is contained in \edits
every other  $(\lambdavec, \kvec)$-lattice.  
 %
%
As mentioned in the introduction for 3D Manhattan sets, it will be
useful to consider the partition of $\mathbb R^d$ induced by the coarse
lattice, whose cells are $k_1 \lambda_1 \times \ldots \times  k_d \lambda_d $
orthotopes (hyper-rectangles) with corners at lattice points.  These orthotopes
will be called \emph{fundamental cells}.  The coarse lattice contains just the
corners of these fundamental cells; other $(\lambdavec,\kvec)$-lattices may
contain points on their edges and faces, but only the dense lattice $L_{\mathbf 1}$ contains points in their interiors.

A third lattice to consider is $L_{\evec_i}$ corresponding to bi-step vector $\bvec = \evec_i $,   
which can be viewed as a collection of points spaced densely on lines
parallel to $\evec_i$, with one line passing through each point of
$(d-1)$-dimensional cubic lattice $L(k_1
\lambda_1,\ldots, k_{i-1} \lambda_{i-1}, k_{i+1} \lambda_{i+1}, \ldots, k_d
\lambda_d)$.  Finally, we mention the lattice corresponding to $\bvec = \mathbf 1 -
\evec_i $, which can be viewed as sampling densely on shifts of the $d-1$
dimensional lattice $L(\lambda_1,\ldots, \lambda_{i-1}, \lambda_{i+1}, \ldots, \lambda_d)$ 
spaced $k_i \lambda_i$ apart.


Given two (binary) bi-step vectors $\bvec_1$ and $\bvec_2$, define their \emph{union} $\bvec_1 \vee \bvec_2 $ and 
\emph{intersection} $\bvec_1 \wedge \bvec_2$ to be their element-wise `OR' and `AND', respectively,
and define $\bvec_1 \subset \bvec_2$ to mean $\bvec_1 \wedge \bvec_2 = \bvec_1$.   
Define the complement to be $\bvec^c \triangleq \mathbf 1 - \bvec$, and the \emph{Hamming weight} or simply \emph{weight}
$\|\bvec\|$ to be the number of ones contained in $\bvec$.

The following are useful properties of  $\gamma$ representations of  
bi-step lattices. 
\vspace{1ex}
\begin{fact}    \label{fact:sublatticeSubsets} 
Considering $(d,\lambdavec, \kvec)$-lattices, 
\begin{description}
\itemsep=3pt

\item[(a)] 
$L_{\bvec_1} \subset L_{\bvec_2}$ 
  if and only if   $\bvec_1 \subset \bvec_2\,,$ 
  
\item[(b)] 
$L_{\bvec_1} = L_{\bvec_2}$
if and only if $\bvec_1 = \bvec_2 \,, $ 

\item[(c)]   $L_{\bvec_1} \cap L_{\bvec_2} ~=~   L_{\bvec_1 \wedge \bvec_2}
  \, ,$ 

\item[(d)]  If  $L_{ \widetilde \bvec} \subset \bigcup_{j=1}^m L_{\bvec_j}$,
then for some $j$, $L_{\widetilde \bvec} \subset L_{\bvec_j}$, 
and consequently from (a), $\widetilde \bvec \subset \bvec_j$.

\end{description}
\end{fact}
%

\vspace{1ex}
\noindent \emph{Proof:}
\vspace{1ex}

\noindent (a) and (b) are elementary.

\vspace{1ex}
\noindent (c)     $L_{\bvec_1} \cap L_{\bvec_2}$
\begin{align*}
        &= \big \{ \tvec :  t_i  \mbox{ is a multiple of $k_i \lambda_i $ for all $i$ s.t.  $b_{1,i} = 0$ or }
            \mbox{$b_{2,i} = 0$, and  $t_i$ is a multiple of $\lambda_i$ for other $i$} \big \} \\
        &= \big \{ \tvec :  t_i  \mbox{ is a multiple of $k_i \lambda_i $ for all $i$ s.t.  $(\bvec_1 \wedge \bvec_2)_i $}
           \mbox{$= 0$ and  $t_i$ is a multiple of $\lambda_i$ for other $i$} \big \} \\
       &= L_{\bvec_1 \wedge \bvec_2}
\end{align*}

\noindent (d)  As is well known, for $m=2$ this property derives from just the group nature of lattices,
but not for larger values of $m$. 
\iftoggle{comment}{\textbf{\textit{ (still need references.  maybe a trip to the math library
is needed to browse algebra books)}}}{}  Accordingly, to prove it for arbitrary $m$,
we need to use properties of $(\lambdavec,\kvec)$ lattices.  Specifically, we demonstrate the contrapositive.
Suppose $L_{\widetilde \bvec} \not \subset L_{\bvec_j}$, $j=1,\ldots,m$.  
Then from (a), for each $1 \leq j \leq m$, $\widetilde \bvec \not \subset  \bvec_j$, 
and so there exists $i_j$ such that $\widetilde b_{i_j} = 1$ and $b_{j,i_j} = 0$.
 Let $I$ denote the set of all such $i_j$'s, and 
let  $\tvec = (t_1, \ldots, t_d)$ be defined by  
\begin{align*}
      t_{i} &= \begin{cases}   (k_i +1) \lambda_i, & \mbox{ if } i \in I \\
                           k_i \lambda_i, & \mbox{ otherwise}
                     \end{cases} .
\end{align*}
Referring to \eqref{eq:Llambdakb}, we see that $\tvec  \in L_{\widetilde \bvec}$, 
because the only dimensions $i$ for which $t_i$ is not a multiple of $k_i \lambda_i$ are those in  $I $, in which
case $\widetilde b_i = 1$, i.e., the lattice $L_{\widetilde \bvec}$ is dense in dimension $i$.
Moreover, again referring to \eqref{eq:Llambdakb}, we see that for each $j \in \{1, \dots, m \}$, 
$\tvec  \not \in L_{\bvec_j}$, 
because 
$t_{i_j} = (k_{i_j} + 1) \lambda_i$  is not a multiple of $k_{i_j} \lambda_i$,
yet $b_{j,i_j} = 0$, i.e., the lattice $L_{\bvec_j}$  is coarse
in dimension $i_j$.
%
%
It follows that 
$\tvec \notin \cup_{j=1}^m L_{\bvec_j}$.  Hence, 
$L_{\widetilde \bvec} \not \subset \cup_{j=1}^m L_{\bvec_j}$.
\hfill $\square$ \edits



\vspace{1.5ex} 
Among other things, (b) shows there is a one-to-one
correspondence between bi-step vectors $\bvec$  and $(d,\lambdavec,
\kvec)$ lattices.  which verifies that the  $\bvec$'s are valid
representations of the $(\lambdavec, \kvec)$-lattices.  Also, since there are
$2^d$ possible bi-step vectors $\bvec$, it follows that there are $2^d$
distinct $(d,\lambdavec, \kvec)$-lattices. 

Note that while (c) shows that the
intersection of two $(d, \lambdavec, \kvec)$-lattices is another $(d,
\lambdavec, \kvec)$-lattice, such is not true for the union, which is why a
Manhattan set is not ordinarily a lattice.  For example, when $d=2$, $
L_{\evec_1} \cup L_{\evec_2} $ is the Manhattan set $ M \big( \big\{ \evec_1, \evec_2
\big\}  \big)$  shown in Fig.\,\ref{fig:twodim}(a), which is not a lattice and
does not equal  $L _{ \evec_1 \vee   \evec_2  }  $, which is the dense lattice
$L_{(1,1)}$.


\subsection{Examples of Manhattan sets} 
\begin{figure*}[!t] \centerline{ 
  \subfloat[]{\includegraphics[width=1.5in]{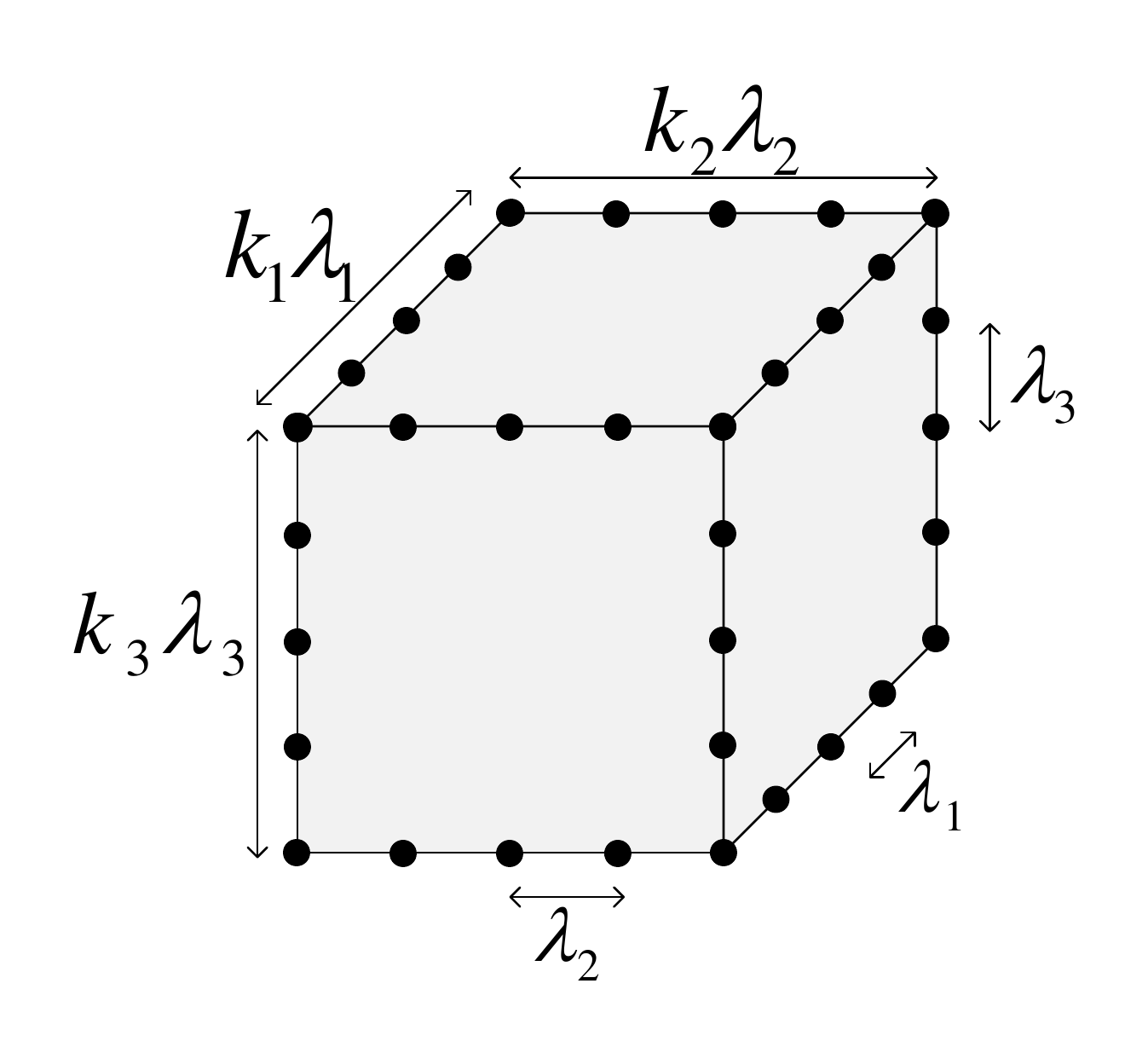}
  }
  \hfil
  \subfloat[]{\includegraphics[width=1.5in]{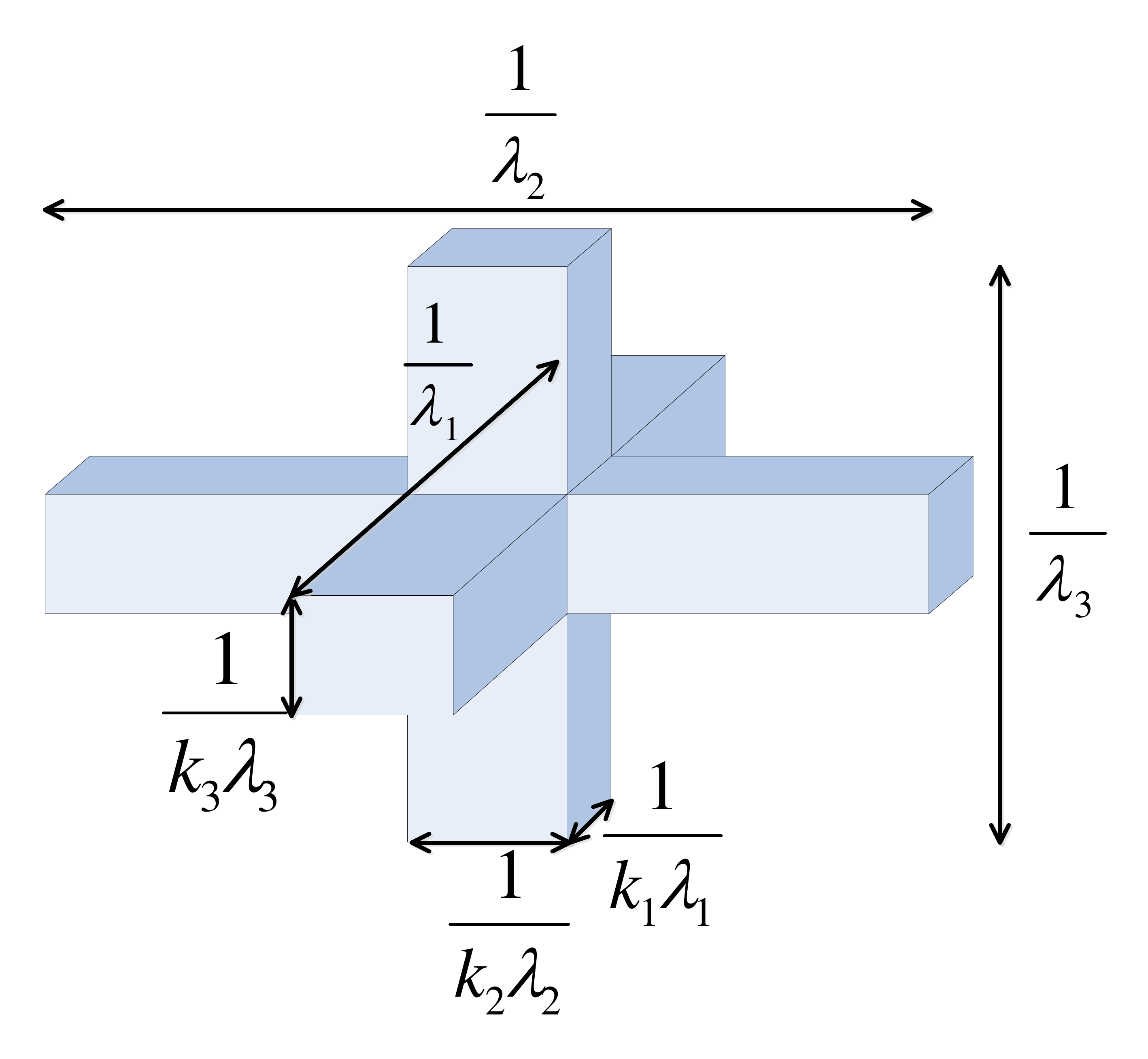}
  } 
  \hfil
  \subfloat[]{\includegraphics[width=1.5in]{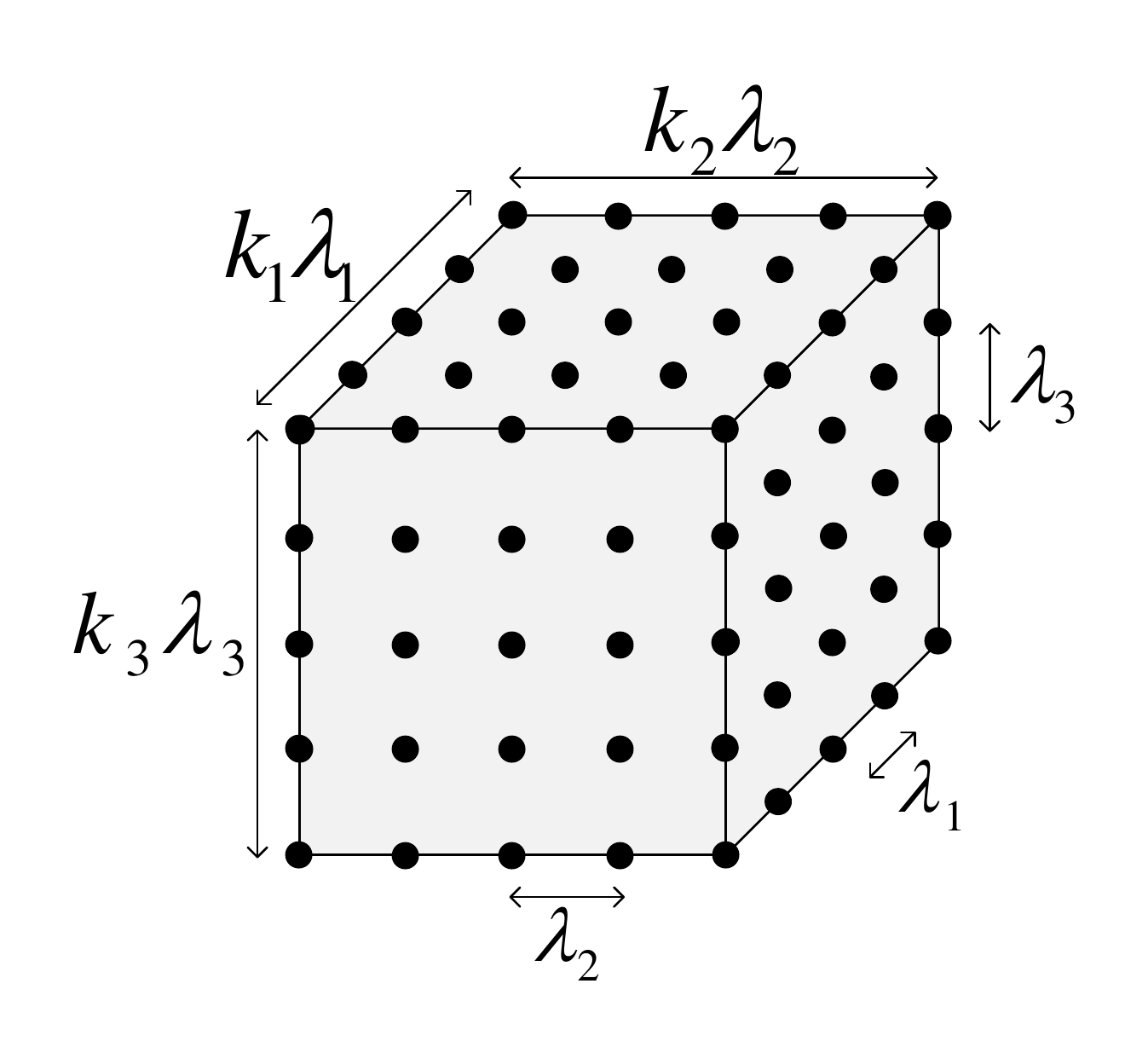}}
  \hfil
  \subfloat[]{\includegraphics[width=1.5in]{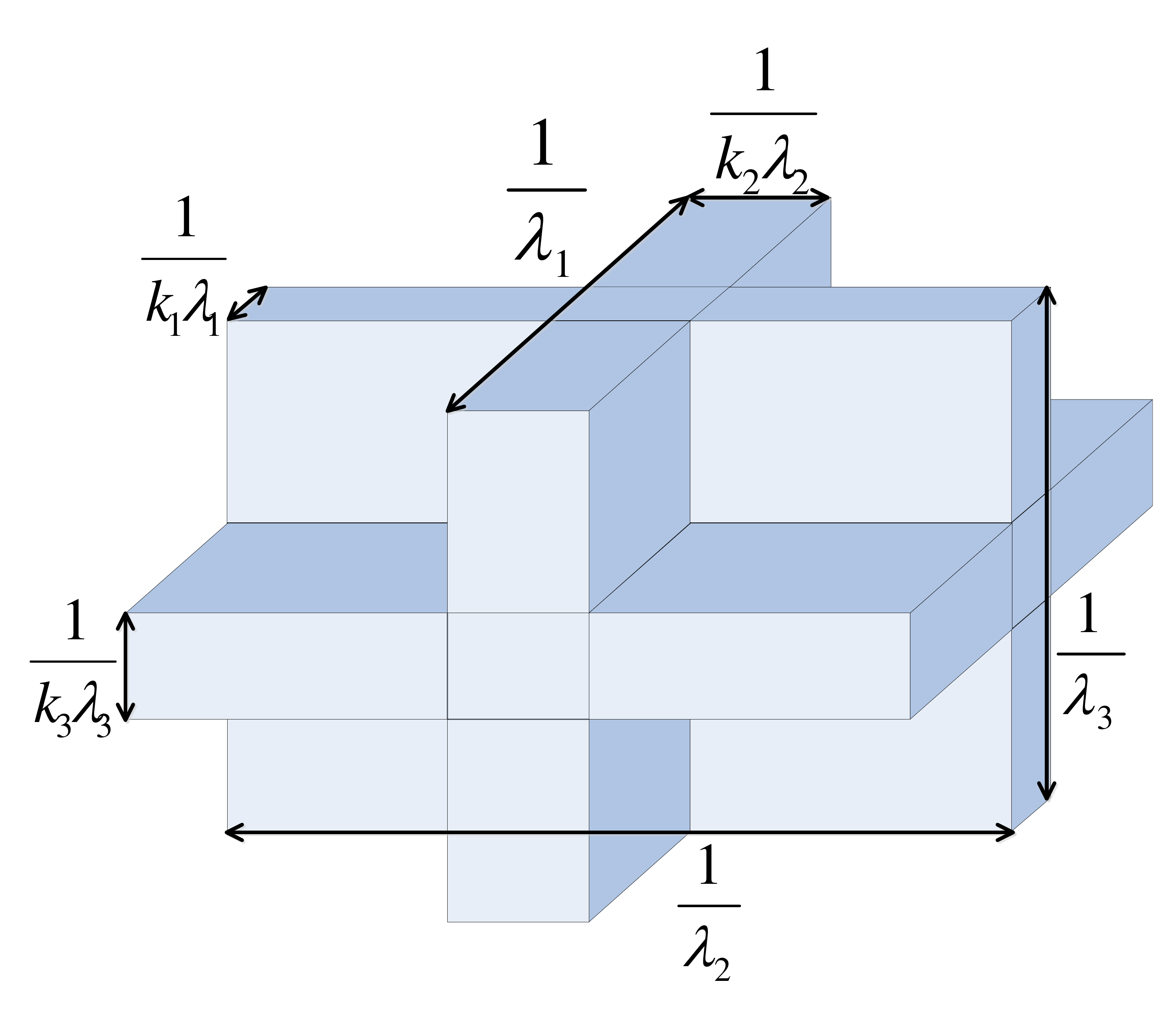}} } 
  \caption{Examples of 3D Manhattan sampling $M(B)$ 
  and their corresponding Manhattan regions $\M(B)$.  
  (a) Manhattan lines $B=\{(1,0,0),(0,1,0),(0,0,1)\}$, and (b) its
  corresponding Manhattan region.  (c) Manhattan facets $B = \{
  (1,1,0),(1,0,1),(0,1,1)\}$, and (d) its corresponding Manhattan region.}
  \label{fig:3d_examples}
\end{figure*}

Several types of Manhattan sets deserve special attention.

\begin{enumerate}

\item  \emph{Manhattan lines}  is a Manhattan set $M(B)$ specified by
  $B = \{ \evec_1,\dots,\evec_d \}$.  
In this case the samples are taken on the 1D edges of the
fundamental cells, i.e., on $d$ orthogonal sets of parallel lines in $\R^d$.
See Figures \ref{fig:twodim}(a) and \ref{fig:3d_examples}(a) for illustrations
of Manhattan lines in two and three dimensions, respectively.

\item  \emph{Manhattan facets} is a Manhattan set $M(B)$ specified by
  $B =  \{ \evec_1^c, \ldots, \evec_d^c  \}$.
  Sampling on a
  set of Manhattan facets is analogous to sampling densely along $d$ orthogonal
  sets of parallel hyperplanes in $\R^d$.  See Figures \ref{fig:twodim}(a)
  and \ref{fig:3d_examples}(c) for illustrations of Manhattan facets in two and
  three dimensions, respectively.  
  For $d=2$,
  Manhattan facets and lines are identical. 

\item Though technically any $(\lambdavec, \kvec)$-lattice, including the
  coarse and dense lattices, is a Manhattan set, we focus on Manhattan sets that are
 not lattices, which we call \emph{proper}. 

\item Video sampling: Let $d=3$, let $i=1,2$ be spatial dimensions and let $i=3$ be a temporal
  dimension.  Whereas video is most commonly sampled with a rectangular lattice, say 
  $L(\lambda_1,\lambda_2,\lambda_3)$,  other samplings are possible, for example,
  the Manhattan set $M(3,\lambdavec, \k, B)$ specified by $B = \{ \evec_3^c,
  \evec_3\}$ uses fine spatial sampling every $k_3 \lambda_3$ seconds
 and spatial subsampling with factors $k_1$ and $k_2$ at times that are
 other multiples
 of $\lambda_3$ seconds.

\end{enumerate}

\subsection{Alternate representations of Manhattan sets}

By the definition of a Manhattan set (\ref{eq:M-grid}) and Fact
\ref{fact:sublatticeSubsets}(a),  augmenting an M-collection $B$ by
a subset  $\bvec'$ of some $\bvec$ in $B$ does not change the resulting
Manhattan set.  Thus many M-collections can generate the same Manhattan set. 
There are, however, unique largest and smallest M-collections
that generate any given Manhattan set.
To find these, we make use of the following. 

\vspace{1ex}

\begin{fact}  \label{fact:Manhattangenerators}

Let  $B$ and 
$B' $ be M-collections.  Then,
\begin{description}

\itemsep=3pt 

\item[~~(a)]  $M(B) \subset M(B')$  if  $B \subset B'$.

\item[~~(b)] $M(B) \subset M(B')$ if and only if for each $\bvec \in B$ there
is $\bvec' \in B'$ such that $L_{\bvec} \subset L_{\bvec'}$, or equivalently
by Fact \ref{fact:sublatticeSubsets}(a),  $\bvec \subset \bvec'$.

\item[~~(c)]  It is not true that  $M(B) = M(B')$ implies
$B = B'$, or that   $M(B) \subset M(B')$ implies $B \subset B'$.

\end{description}
\end{fact}

\vspace{1ex}
\noindent \emph{Proof:}

\iftoggle{comment}{\footnote{temp:  we agree not to move this proof to the appendix 
because it is  very short and there's nothing else to put there.}}{}
(a) is obvious.  

\vspace{.5ex}
\noindent (b)   If  for each $\bvec \in B $ there
is $\bvec' \in B'$ s.t. $L_{\bvec} \subset L_{\bvec'}$,
then  $M(B) = \cup_{\bvec \in B} L_{\bvec} \subset 
\cup_{\bvec' \in B'} L_{\bvec'} = M(B')$.
Conversely, if $M(B) \subset M(B')$, then for each $\bvec \in B$,
$L_{\bvec} \subset M(B') = \cup_{b' \in B'} L_{\bvec'}$, 
and Fact \ref{fact:sublatticeSubsets}(d) implies $L_{\bvec} \subset
L_{\bvec'}$ for some $\bvec' \in B'$.

\vspace{.5ex}
\noindent (c)
Part (a) shows that adding to $B$ some subset of  some $\bvec \in B$
that is not already in $B$ yields $B' \not = B$ such that $M(B) = M(B')$.
In this same case,  $M(B') \subset M(B)$, but
$B' \not \subset B$.
\hfill $\square$

\vspace{1ex}
The unique largest M-collection that generates $M(B)$ is
\begin{align*}
    \overline B ~\triangleq~  \big \{\bvec' : \bvec'  \subset \bvec \mbox{ for some } \bvec \in B \big \} \, ,
\end{align*}
which 
we call the \emph{closure} of $B$. 
Fact \ref{fact:Manhattangenerators}(a) implies $M(B) \subset M(\overline B) = M(B)$, and 
Fact \ref{fact:Manhattangenerators}(b) implies $M(\overline B)  \subset M(B)$.  Hence $M(\overline B) = M(B)$.
$M(\overline B)$ is the largest M-collection generating 
$M(B)$ because if  $M(B') $ $= M(B)$, and $\bvec' \in B'$, then Fact \ref{fact:Manhattangenerators}(b) implies there is a $\bvec \in B$ such that  $L_{\bvec'} \subset L_\bvec $, and this implies $\bvec' \in \overline B$.   Hence, $B' \subset \overline B$.

Removing all elements of an M-collection $B$ that are subsets of another element results in the unique smallest M-collection generating $M(B)$, which we denote $\underline B$.   

\subsection{Manhattan sampling density}

The \emph{density} of a Manhattan set $M(B)$, i.e., the number of samples
per unit area, is obviously less than the sum of the densities of the lattices
of which it is the union, because 
each lattice
contains all points in the coarse lattice $L_{\mathbf 0}$.  
Accordingly, we
partition the dense lattice $L_{\mathbf 1}$ in such a way that for any
$B$,  $M(B)$ is the union of atoms of this partition, and its density
can be computed by summing their densities.
 
\begin{figure}[!t] 
 \centering
 \parbox[t]{3in}{%

	\centering
	 \includegraphics[width=2in]{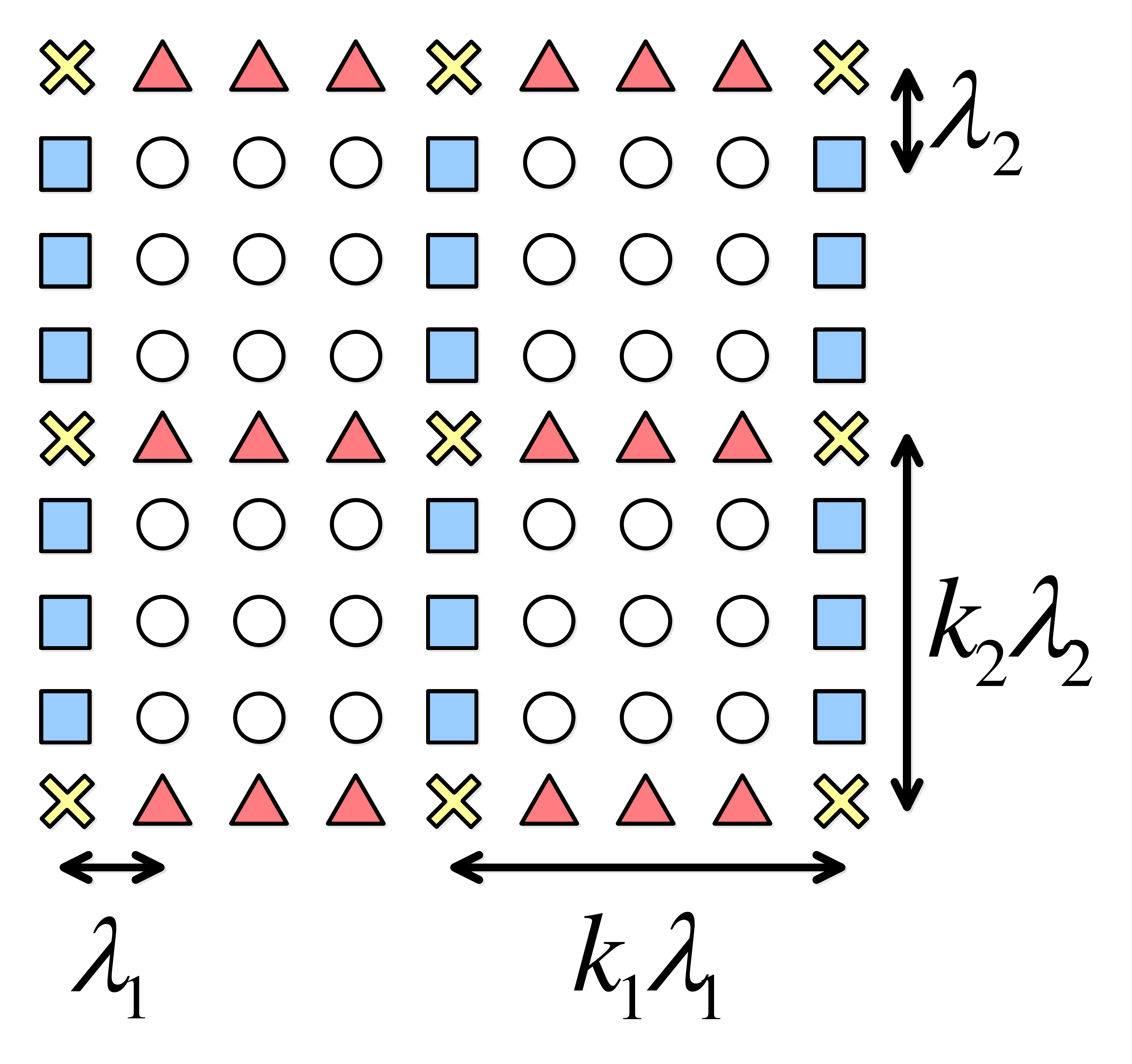}

  \vspace{-2ex} \caption{Partitioning of 2D Manhattan grid 
  $M(\{ \evec_1,\evec_2 \} ) $ with
  sampling factors $k_1 = k_2 = 4$, which is the union of the yellow $\times$'s in
  $V_{(0,0)}$, the red $\triangle$'s in $V_{(1,0)}$, and the blue
  $\square$'s  in $V_{(0,2)}$.  The white $\circ$'s are in 
  $V_{(1,1)}$, which is disjoint from  this Manhattan grid. }
  \label{fig:partitionedGrid} 
  }
 ~\,
\parbox[t]{3in}{%

 \centering  
 \includegraphics[width=2.5in]{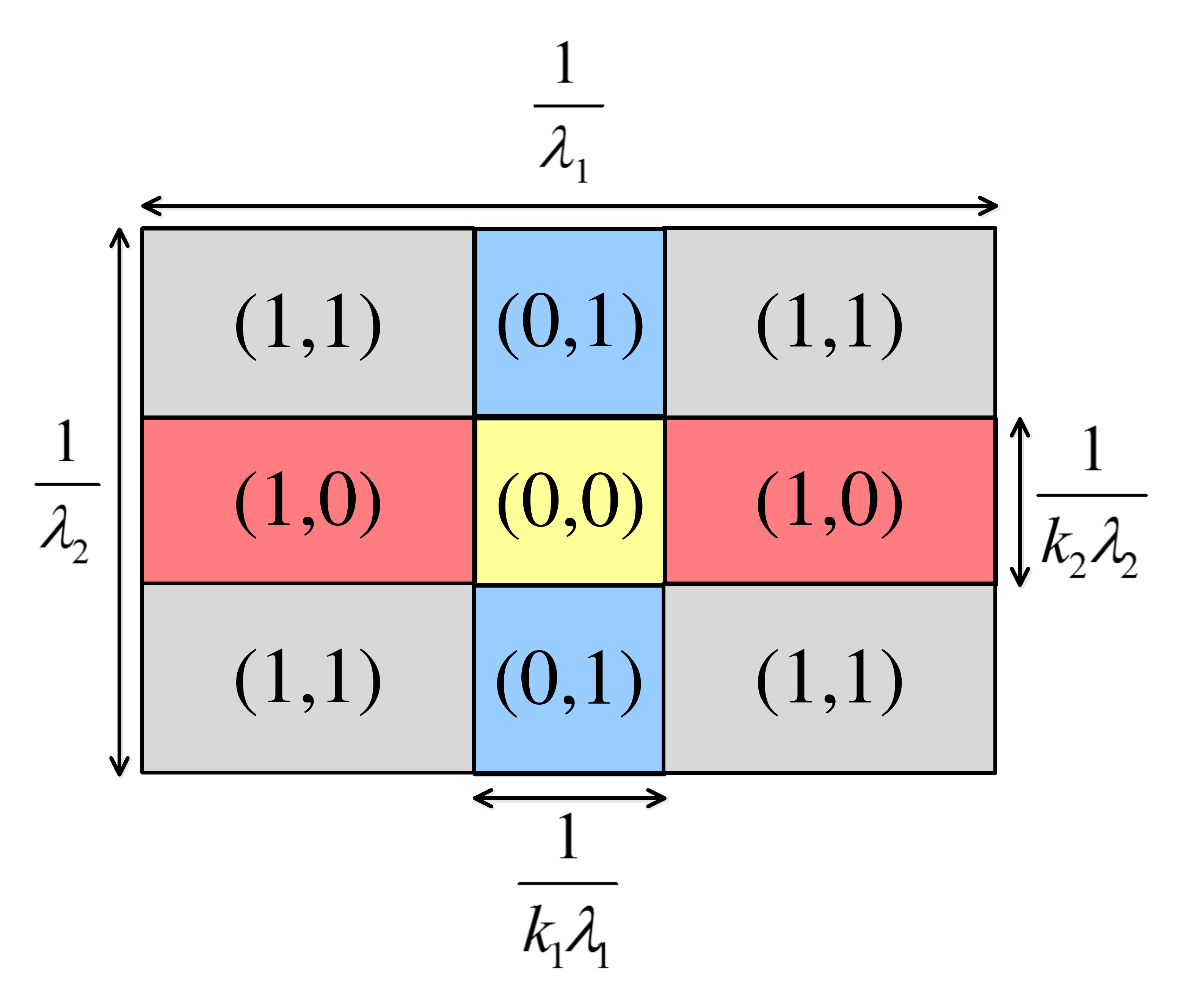}
  \vspace{-2.02ex} \caption{
  M-partition of $\mathcal N_{\mathbb D}$ for
  $d=2$, $k_1 =5$, $k_2=3$.  Frequency $\u = 0$ lies at the center.  Each
  M-atom $A^\bvec$ is identified by its $\bvec$.  Note that the
 cross-shaped Manhattan region $\M( \{ \evec_1,\evec_2 \} )$ is also partitioned by
  M-atoms; in particular, $\M(\{ \evec_1,\evec_2 \}) = A^{(0,0)} \cup
  A^{(1,0)} \cup A^{(0,1)}$.  }
  \label{fig:partitionedFrequency}
 } 
\end{figure}

To obtain a suitable partition, let us group into one atom all sites $\t$ 
of $L_\onevec$ having the 
same answers to the following $d$ questions -- ``Is $t_i$ not a multiple of 
$\lambda_i k_i$?" -- for $i=1, \ldots, d$.  
Specifically, with a binary vector $\bvec = (b_1, \ldots, b_d)$
indicating the set of $i$'s for which the answers are ``yes", consider the partition
%
%
$ \big \{ V_{\bvec} :  \bvec \subset  \{ 0, 1 \}^d   \big\} $,
  where the atom corresponding to $\bvec$ is 
\begin{align*}        
V_{\bvec} &~\triangleq~  \big \{ \tvec :  \mbox{ $t_i$ is a multiple of $k_i \lambda_i$ for $i$ s.t. $b_i =0$,} \notag \\
        &  \hspace{5ex}  \mbox{ and $t_i$ is a multiple of $\lambda_i$, but not  $k_i \lambda_i$, for $i$ s.t. $b_i=1$} \big\}  
\end{align*}
Fig.\,\ref{fig:partitionedGrid}
illustrates the partitioning of a 2D Manhattan grid.  
Essentially, it is a partition of the dense lattice into
collections of cosets
of the coarse lattice.  

\edits
%

It is clear from the definition that
no $\tvec$ can lie in both $V_{\bvec}$ and $ V_{\bvec'}$ for $\bvec \not=\bvec'$.  
Hence, the $V_{\bvec}$'s are disjoint.
By comparing the above to the definition \eqref{eq:Llambdakb} of $L_\bvec$, one sees
that $V_{\bvec'} \subset L_{\bvec}$ if and only if $\bvec' \subset \bvec$.
It follows that for any $\bvec$,  $\cup_{\bvec' \subset b} V_{\bvec'} \subset L_\bvec$.  
Conversely, if $\t \in L_\bvec$, then it is easily seen that  $\t \in L_{\bvec'}$ for 
$\bvec'$ defined by $b'_i = 0$ for $i$ such that $t_i$ is a multiple of $\lambda_i k_i$
and $b'_i =1$ otherwise.  
It follows that $\cup_{\bvec' \subset b} V_{\bvec'} =  L_\bvec$, 
i.e., $\{ V_\bvec \}$ partitions any bi-step lattice, including $L_{\mathbf 1}$. 
Moreover, since $M(B) = \cup_{\bvec \in B} L_\bvec$, one can also write 
$M(B) = \cup_{\bvec \in \overline B}  V_{\bvec}$, i.e.,
$\{ V_\bvec \}$ partitions any Manhattan grid.


With this partition in mind, the density of the Manhattan set $M(B)$ is
now obtained by summing the densities of each $V_\bvec$, $\bvec \in \overline
B$.  Consider the points of $V_\bvec$ in the  
fundamental cell 
\begin{align*} 
   F_{\kvec,  \lambdavec} ~\triangleq~  \bigtimes_{i=1}^d [0,
  k_i \lambda_i) \, ,  
\end{align*}
which has a corner at the origin, lies entirely in the 
positive hyper-quadrant,  and has volume is $\prod_{i=1}^d k_i \lambda_i$.
  We see that $V_\bvec \cap
F_{\kvec , \lambdavec}$ is the Cartesian product of $d$ sets
$A_1, \ldots, A_d$, where 
$A_i = \{ \lambda_i, 2 \lambda_i,  \ldots, (k_i - 1) \lambda_i \big \} $ 
if $b_i =1$ and $A_i = \{0\}$ if $b_i =  0 $. 
 Since $V_{\bvec} \cap F_{\kvec, \lambdavec}$ contains $\prod_{i: b_i = 1}
  (k_i-1)$  points, 
  and since the density of $V_\bvec$ is this  number  divided
  by the volume of $F_{\kvec,  \lambdavec}$, the density of $M(B)$ is
\begin{align}   \label{eq:rhoGamma} 
     \rho(B) ~=~   {  \sum_{\bvec \in
    \overline B} \, \prod_{i : b_i = 1} (k_i-1)  \over   \prod_{i=1}^d k_i
    }  \times  {1 \over \prod_{i=1}^d  \lambda_i } \, .  
\end{align}
For example, the densities of several Manhattan sets in three dimensions are given in Table
\ref{tab:density}.
\begin{table}[h]
  %
  \centering
  \begin{tabular}{cc|c}
    Manhattan set & $\Gamma$ & $\rho(\Gamma)$\\[.5ex] \hline
    \rule{0pt}{9pt} Manhattan lines & $\{\evec_1,\evec_2,\evec_3\}$ & $(3k-2)/k^3$\\[1ex]
    Video sampling example & $\{\evec_3^c,\evec_3\}$ & $(k^2+k-1)/k^3$\\[1ex]
    Manhattan facets & $\{\evec_1^c,\evec_2^c,\evec_3^c\}$ & $(3k^2 - 3k+1)/k^3$
  \end{tabular}
  \vspace{2.5ex}
  \caption{Density of several 3-dimensional Manhattan sets with $k_i=k$ and
  $\lambda_i=1$ for all $i$. 
  \iftoggle{comment}{{\bf vertical space added in table and before caption}}{} }
  \label{tab:density}
\end{table}

\subsection{Manhattan partition of frequency space}
\label{sec:frequencySpace}


As mentioned earlier,
our approach to reconstructing an
appropriately bandlimited image $x(\tvec)$ from M-samples 
$M(B)$ involves sequentially reconstructing regions of its spectrum
$X(\u)$.  Specifically, each region will be recovered from 
the samples in some collection of bi-step lattices contained
in the Manhattan set.
This section
describes a partition of frequency space, some of whose atoms will be the
reconstructable regions.

Let  $\mathcal N_\bvec$ denote the Nyquist region for bi-step lattice
$L_\bvec$, i.e.,
\begin{align*}  
    \mathcal N_\bvec  ~\triangleq~   \Big \{ \u : | u_i | < {1 \over 2
    \alpha_{\bvec,i}} , \, i =1, \ldots, d \Big\} \, , 
\end{align*}
with step sizes $\alpha_{\bvec,i}$  given by (\ref{eq:alphagamma}).
For future reference we note that 
\begin{align}   \label{eq:nyquistsize} 
  \mathcal N_{\bvec'} \subset \mathcal N_{\bvec}       \mbox{ if and
  only if }  \bvec' \subset \bvec \, .  
\end{align}

Since any $(d,\lambdavec, \kvec)$ Manhattan set is a subset
of the dense lattice $L_{\mathbf 1}$,  
it follows that the appropriate bandlimitation
for images reconstructable from any such Manhattan set 
or any $(d, \lambdavec, \kvec)$ lattice is a subset of the Nyquist region 
of the  dense lattice, namely, 
\begin{align*}
   \mathcal N_{\mathbf 1} ~=~  \Big \{ \u :  | \u_i | < {1 \over 2 \lambda_i} , 
     \, i = 1, \ldots, d \Big \} \, .
\end{align*} 
Thus, we need only partition $\mathcal N_\onevec$.

\vspace{1ex} \begin{definition} The \emph{Manhattan partition} (\emph{M-partition}) of $\mathcal
  N_{\mathbf 1}$  is
$ \big\{  \mathcal A^\bvec : \bvec \in \{0,1\}^d   \big\} $
where  $\A^\bvec$ is the \emph{Manhattan atom}\footnote{Consistent with
previous conventions, $\bvec$ is a
superscript because it determines a frequency region (an M-atom $\A^\bvec$),
whereas it is a subscript when specifying the Nyquist region $\mathcal N_\bvec$ of
a bi-step lattice $L_\bvec$,  precisely because it prescribes a sampling.
Except for $\bvec = \zerovec$, 
the M-atoms are highpass regions, whereas Nyquist regions are lowpass.
}
 \emph{(M-atom})
   \begin{equation*} \A^\bvec  ~\triangleq~   a^\bvec_{1} \times \dots \times
     a^\bvec_{d} \, , \label{eq:criticalRegion} \end{equation*}
  and $a^{\bvec}_i$ is  the interval, or union of two intervals, 
 \begin{equation*} 
     a^{\bvec}_i \,\triangleq \, 
     \begin{cases} \big\{ u_i :
        \frac{1}{2 k_i \lambda_i} \leq |u_i | <  \frac{1}{ 2\lambda_i} \big)  \big\},
        & b_i = 1\\[.75ex] \big\{ u_i : |u_i| < \frac{1}{2k_i\lambda_i}
        \big\}, & b_i = 0
     \end{cases} .  \label{eq:criticalInterval}
 \end{equation*}
 Thus, $\A^\bvec$ is highpass for all dimensions such that $b_i=1$
 and lowpass for all other dimensions.  The \emph{weight} of atom $\mathcal A^\bvec$
 is  $\| \bvec \|$, the weight of $\bvec$. 
\end{definition}

\vspace{1ex}
The M-partition is illustrated in Fig.\,\ref{fig:partitionedFrequency}
in the case of $d=2$, $k_1=5$, $k_2= 3$.
We now make several easy to deduce, but important, observations.

\vspace{.5ex}
\begin{enumerate}
\itemsep = 3 pt 
 
 \item 
M-atom $\A^\bvec$ is the Cartesian product of 
lowpass intervals $(-{1 \over 2k_i \lambda_i}, {1 \over 2k_i \lambda_i})$  
along each dimension such that $b_i=0$, and of the
union of two highpass intervals $(-{1 \over 2 \lambda_i}, -{1 \over 2k_i \lambda_i}]
\cup [{1 \over 2 k_i \lambda_i}, {1 \over 2 \lambda_i})$ along each
dimension such that $b_i =1$.   Thus, $\A^\bvec$ is the union of
$2^{\|\bvec\|}$ disjoint orthotopes in $\R^d$.

\item The M-atoms are disjoint, and
 their union is $\mathcal N_{\mathbf 1}$.  Hence, they comprise 
 a partition of $\mathcal N_{\mathbf 1}$.
 
 \item The M-atom $\A^\bvec$ is a subset of the Nyquist region
	$\mathcal N_\bvec$.  
	Equality holds only for  $\bvec = \zerovec$.
	
\item The weight $\|\bvec\|$ of an atom $\A^\bvec$ is a rough indicator of how highpass
or lowpass is the atom.
	
 \item The lowest weight M-atom, $\A^\zerovec$, is lowpass in all dimensions 
 and equals  the Nyquist region
 $\mathcal N_\zerovec$  of the coarse lattice
	$L_\zerovec$, which is the bi-step lattice with smallest and lowest 
	frequency Nyquist region. 

  \item The highest weight M-atom, $\A^{\mathbf 1} $, is highpass in all dimensions and
    contains the highpass ``corners'' of $\mathcal N_{\mathbf 1}$.  
    Its volume is at least large as that of any other atom, and usually larger.
    We will see later that no proper $(d, \lambdavec, \kvec)$ Manhattan set
    permits the reconstruction of these corners since they can only be
    recovered by sampling densely along every dimension, i.e., they are only
    recoverable if we sample on the dense lattice $L_{\mathbf 1}$.  


  \item  If an 
  image $x(\t)$ is bandlimited to $\Ncal_\onevec$, then both $x(\t)$ and its spectrum
  $X(\u)$ can be decomposed into sums of 
  M-atom components:
  \begin{align} 
    x(\t) &= \sum_{\bvec \in \{0,1\}^d}  x^\bvec(\t) 
          \nonumber  \\[.5ex]
          X(\u) &=  \sum_{\bvec \in \{0,1\}^d } X^{\bvec}(\u) \, ,
          \label{eq:Xdecomposegamma}
   \end{align} 
  where $X^\bvec(\u) = X(\u)$ for $\u \in \A^\bvec$, 
  $X^\bvec(\u) = 0$ otherwise, and $x^\bvec(\t)$ is the inverse
  transform of $X^\bvec(\u)$.  
  We shall refer to $x^b(\t)$ and $X^\bvec(\u)$ as \emph{Manhattan atoms},
  or simply \emph{atoms}, 
  of $x(\t)$ and $X(\u)$, respectively.

\end{enumerate}

\vspace{1ex}
It will also be important that the M-atoms can partition
the Nyquist region  $\mathcal N_\bvec$ corresponding to any 
bi-step lattice $L_\bvec$, as shown below. 



\vspace{1ex}
\begin{fact}   \label{fact:partitionProperty} 
~\\[-2ex]
\begin{description}
\itemsep=3pt

\item[(a)] 
  For any $\bvec \in \{0,1\}^d $, 
the $2^{\|\bvec\|}$ M-atoms in  
$\{ \A^{\bvec'} :  \bvec' \subset \bvec \}$
 partition $\mathcal N _\bvec$ in the sense that
 $\mathcal N_\bvec = \cup_{\bvec' \subset \bvec} \A^{\bvec'}$.  
 
\item[(b)]  
$\bvec' \subset \bvec$ if and only if  $\A^{\bvec'}  
\subset \mathcal N_\bvec$.

\end{description}
\end{fact}

\vspace{1.5ex}
\noindent \emph{Proof:}   \edits

\iftoggle{comment}{\footnote{temp: again, we decided not to move this proof
to the appendix}}{}
%
 (a)  Since the elements of $\{ \A^{\bvec'} :  \bvec' \subset \bvec \}$ 
 are disjoint and contained in $\mathcal N_\bvec$,
 it suffices to show 
\begin{equation}   \label{eq:unionOfAllCriticalRegions}
     \mathcal N_\bvec ~\subset~  \bigcup_{\bvec^\prime \subset \bvec}
	\A^{\bvec^\prime}  \, .
 \end{equation}
Accordingly, suppose $\u \in \mathcal N_\bvec$.
%
 %
It is then easy to see that  $\u \in \A^{\bvec'}$, where 
%
\[b_i' = 
\begin{cases}
  1, & {1 \over 2k_i \lambda_i}  \leq |u_i| < {1 \over 2 \lambda_i} \\
  0, & |u_i| \leq {1 \over 2k_i \lambda_i} 
\end{cases},
\]
%
which demonstrates (\ref{eq:unionOfAllCriticalRegions}).

(b)  First, if  $\bvec' \subset \bvec$, then by (\ref{eq:nyquistsize}) 
and the third observation after the definition of M-atom,  
$\A^{\bvec'} \subset \mathcal N_{\bvec'} \subset \mathcal N_\bvec$.
Conversely, if $\A^{\bvec'} \subset \mathcal N_\bvec$, then by part (a)  $\A^{\bvec'}$ must be one of the atoms whose union is $\mathcal N_\bvec$.
Hence,
$\bvec' \subset \bvec$.
$\hfill \square$

\subsection{Spectral replication induced by bi-step lattice sampling}
\label{sec:spectralReplications}


%
%
 
For a Manhattan set  $M(B)$, the reconstruction algorithm to follow
will reconstruct an image $x(\t)$ by reconstructing its spectrum $X(\u)$
 one Manhattan atom at a time, in an order to be specified later.  
 In particular, for each $\bvec \in B$,  it will reconstruct atom
 $X^\bvec(\u)$ from the subset of samples corresponding to the bi-step lattice $L_\bvec$,
 taking  into account the aliasing due to previously reconstructed atoms.   
Using the more suggestive $\svec = (s_1, \ldots, s_d)$, rather than $\bvec$, 
to denote a bi-step vector that characterizes a sampling, then as reviewed in Section \ref{sec:prelims}, 
sampling with $L_\svec$  replicates the spectrum 
$X(\u)$ at all sites in the reciprocal lattice $L^*_{\svec}$.
Moreover, if $X(\u)$ is bandlimited to $\Ncal_\onevec$, substituting \eqref{eq:Xdecomposegamma}
 into \eqref{eq:sampledspectrum} gives the following decomposition 
 of the sampled spectrum:  
  \begin{align} 
         X_\s(\u) ~=~   \sum_{\v \in L^*_\svec}   \, \sum_{\bvec \in \{0,1\}^d }   X^{\bvec}(\u-\v) \, .
          \label{eq:Xdecomposegamma2}
 \end{align} 
 We will refer to the term $X^\bvec (\u -\v)$, 
and its spectral support  $\A^\bvec + \v  \triangleq \{ \u + \v: \u \in \A^{\bvec} \}$, 
as the replica of atom $X^\bvec(\u)$, respectively, $\A^\bvec$, at site $\v$. 
Using this terminology, one sees that reconstructing atom $X^\svec(\u)$
from the sampled spectrum $X_\svec(\u)$ requires accounting for the potential aliasing, i.e., overlap,
of the replicas of the various atoms of $X(\u)$
on $X^\svec(\u)$.  
This requires knowing
which replicas of each atom
will alias, i.e. overlap, $\A^{\svec}$.  More specifically,
since the algorithm will only apply to images whose spectral support is limited 
to some subset of the Manhattan atoms, for any pair of bi-step vectors $\bvec$ and $\bvec'$,
we will need to know whether  sampling with bi-step lattice $L_\svec$ causes 
 a replica of atom $\A^{\bvec'}$
(at some site $\v \in L^*_{\svec}$) to overlap atom $\A^\bvec$ of the original spectrum. \edits

%


%

Such overlap questions are answered by the following  lemma and its corollary.
Let  $R^{\bvec'}_{\svec}$ denote the union of 
the replicas of all $\bvec'$ atoms
induced by sampling with $L_{\svec}$.
That is, 
\begin{align*}    
	R^{\bvec'}_{\svec}  
    %
    %
    %
	    ~\triangleq~      \bigcup_{\v \in L^{^*}_{\svec} - \ovec }
	                 \big[ \A^{\bvec'} + \v \big]    
    %
        \, .
\end{align*}
%

\vspace{1ex}

\begin{lemma}   \label{lemma:SOmegaOverlap}
Consider sampling with $L_{\svec}$. \\[-2ex]
\begin{description}
\itemsep=3pt

\item[(a)]
For all  $\bvec, \bvec' \subset \svec$, no 
replica of $\A^{\bvec'}$
 overlaps $\A^{\bvec}$, i.e.,
 $R_{\svec}^{\bvec'} \cap \A^{\bvec} = \emptyset$. 


\item[(b)] 
The replicas of $\A^{\bvec'}$ 
 induced by sampling with $L_{\svec}$ do not overlap $\A^{\bvec}$ 
 if 
there exists at least one dimension $i$ 
such that $s_i = 1$ and 
$b_i \neq b'_i$.
That is, $R^{\bvec'}_{\svec}\cap \A^{\bvec} = \emptyset$ if 
$(\bvec \oplus \bvec' ) \wedge \svec \neq \zerovec$,
where $\bvec \oplus \bvec'$ denotes element-wise exclusive or (XOR). 

\end{description}
\end{lemma}

\vspace{1ex}
\emph{Proof:}

 (a)    If  $\bvec, \bvec' \subset \svec$, then Fact
 \ref{fact:partitionProperty}(b) shows $\A^\bvec, \A^{\bvec'} \subset
 \mathcal N_{\svec}$.  Since the sampling theorem for conventional
 rectangular lattice sampling shows that replicas of $\mathcal N_{\svec}$ do
 not overlap $\mathcal N_{\svec}$, it follows that the replicas of
 $\A^{\bvec'}$ cannot overlap $\A^{\bvec}$.


(b)  Let us  compare  the M-atom 
\begin{align*}
     \A^{\bvec}   ~=~   a^\bvec_{1} \times \dots \times a^\bvec_{d} 
\end{align*}
to an arbitrary replica in $R^{\bvec'}_{\svec} $:
\begin{align*}
     \A^{\bvec'} + \v  ~=~   (a^{\bvec'}_{1} + v_1) \times \dots \times 
     (a^{\bvec'}_{d} + v_d ) \, ,
\end{align*}
for  $\v \in L^*_{\svec} - \{ \bf 0 \} $.   Note that
$\A^{\bvec}$ and $\A^{\bvec'} + \v$ are disjoint if and only if
$a^\bvec_{i}  \cap ( a^{\bvec'}_{i} +  v_i) = \emptyset $, for some $i$.


If, as hypothesized in the lemma,    $(\bvec \oplus \bvec' )\wedge \svec \neq \zerovec$.
Then there must  exist $i$ such that $s_i =1 $ and either
$b_i=1$, $b'_i=0$ or $b_i=0$, $b'_i=1$.
First, consider the case that $b_i=1$, $b'_i=0$.  Then 
\begin{align*}
   a^\bvec_{i} ~=~ \Big \{ u_i : {1 \over 2 k_i \lambda_i }
                           \leq | u_i | < { 1 \over 2 \lambda_i } \Big \}
\end{align*}
and
\begin{align*}
a^{\bvec'}_{i} + v_i 
    &~=~ \Big \{ u_i :  |  u_i  | <  { 1 \over 2 k_i  \lambda_i } \Big \} + v_i \, .
\end{align*}
Since $s_i = 1$, we have 
{$v_i = {n_i \over \lambda_i}$ for some $n_i$, and one sees 
from the above that no matter the value of $n_i$,  $(a^{\bvec'}_i + v_i) 
\, \cap \, a^{\bvec}_i = \emptyset$.  Hence,  
$(\A^{\bvec'} + \v)  \cap \A^{\bvec} = \emptyset$, and so
$R^{\bvec'}_{\svec} \cap \A^{\bvec} = \emptyset$.
A similar argument applies for the case that $b_i=0$, $b'_i=1$. 
\hfill $\square$

\vspace{1ex}

The following will provide a key step in showing how to reconstruct 
appropriately bandlimited images.

\vspace{1ex}
\begin{corollary}      \label{cor:biggestGamma} 
If $\| \bvec' \| \leq \| \svec \|$, 
then replicas in  $R^{\bvec'}_{\svec}$ do not overlap $\A^{\svec}$.   
\end{corollary}
\vspace{1ex}
\emph{Proof:}
We will apply Lemma \ref{lemma:SOmegaOverlap} with $\bvec = \svec$.
If $\bvec'=\svec$, then Part (a) of Lemma \ref{lemma:SOmegaOverlap} shows $R^{\bvec'}_{\svec} 
\cap \A^{\svec} = \emptyset$.  
If $\bvec' \neq \svec$ and $\|\bvec'\| \leq
  \|\svec\|$, then there must exist $i$ such that $s_i =1$ and $b'_i=0$.
Therefore, 
$(\svec \oplus \bvec') \wedge \svec \neq \zerovec$, and Part (b) of
Lemma \ref{lemma:SOmegaOverlap} shows $R^{\bvec'}_{\svec} 
\cap \A^{\svec} = \emptyset$. 
\hfill $\square$

\subsection{The multidimensional Manhattan sampling theorem}

Given a Manhattan set $M(B)$, consider its \emph{Manhattan region},
which is defined to be the union
of the Nyquist regions of the bi-step lattices of which it is the 
union:
\begin{align*}   	\label{eq:recoverableManhattanRegion}
    \mathcal M(B) &~\triangleq~ \bigcup_{\bvec \in B} \mathcal N_\bvec
	 ~=~  \bigcup_{\bvec \in B} \;  \bigcup_{\bvec' \subset \bvec} 
	         \A^{\bvec'} \nonumber 
   ~=~  \bigcup_{\bvec^\prime \in \overline B} A^{\bvec^\prime}  ,
\end{align*}
where the second equality uses Fact \ref{fact:partitionProperty}(a).
In this section we show that images bandlimited
to  $\mathcal M(B)$ can be recovered
from their M-samples in $M(B)$;  we give an
explicit procedure for reconstructing such images
from their samples in $M(B)$; and we show that the set of such
bandlimited images is maximal in the Landau sense.

The key steps are the next two lemmas.  The first shows that for any image
$x(\tvec)$ whose spectrum $X(\u)$ is bandlimited to $\mathcal M(B)$, 
the portion  of $X(\u)$ in any highest weight M-atom
$\A^\bvec$ can be easily recovered from the samples in
$L_\bvec$, which are a subset of the M-samples.  Specifically,
$X^\bvec(\u)$ can be 
recovered simply by extracting the $\A^\bvec$
portion of the sampled spectrum $X_\bvec(\u)$  due to sampling with $L_\bvec$.
Equivalently, the corresponding component $x^\bvec(\t)$ of $x(\t)$ can be
recovered by filtering the sampled image $x_\bvec(\t)$ with an ideal
bandpass filter with frequency support $\A^\bvec$.

\vspace{1ex}

\vspace{1ex} \begin{lemma}   \label{lem:highestOmega} 
Suppose $M(B)$ is a
  Manhattan set and $x(\t)$ is an 
  image whose spectrum $X(\u)$
  is bandlimited to $\mathcal M(B)$.  
  Then if   $\bvec$ has maximal weight in
  $\overline B$,
\begin{equation}   	  \label{eq:highestOmega} 
    X^\bvec(\u) ~=~ H^\bvec(\u) \, X_\bvec (\u)  \, , 
\end{equation}
where    
$H^\bvec(\u) = 1$ for $\u \in A^\bvec$, and 0 otherwise.
\end{lemma}

\vspace{1ex}

\emph{Proof:}
 Consider any $\bvec$ of maximal weight.
 Since, according to (\ref{eq:sampledspectrum}), $X_\bvec(\u)$ consists of replicas of $X(\u)$ at the frequencies in $L^*_\bvec$, since $X(\u)$
 can be decomposed into its components on Nyquist
 atoms $\{ \A^{\bvec'}: \bvec' \in  \{0,1\}^d \}$,  
 and since $X(\u)$ is bandlimited to 
 $\mathcal M(B) = \bigcup_{\bvec' \in \overline B} \A^{\bvec'}$, 
 it suffices to argue that for all $\bvec' \in \overline B$, 
 no replica of $\A^{\bvec'}$ intersects $\A^\bvec$.  
 First, Lemma \ref{lemma:SOmegaOverlap}(a) applied with $\svec =  \bvec' = \bvec$ 
 shows that no replica of
 $\A^\bvec$ with $\v \neq {\bf 0}$ intersects $\A^\bvec$.  Second, Corollary \ref{cor:biggestGamma}
 and the fact that $\bvec$  has maximal weight in $\overline B$ 
 imply that for any other $\bvec' \in \overline B$, 
 no replica of $\A^{\bvec'}$ can overlap $\A^\bvec$.  
 \hfill $\square$
 
 \vspace{1ex} 

Once $X(\u)$ has been
recovered in all such highest weight (highest frequency) M-atoms, the next lemma shows
that $X(\u)$ can then be recovered in the next highest weight M-atoms by canceling the contributions
due to atoms with larger weight.  In effect,
the aliasing of one atom comes only from atoms with larger weight, i.e.,
higher frequency.
\edits
%
%
%
Specifically, for any such $\bvec$, it shows that 
$X^\bvec(\u)$ can be recovered from the spectrum
$X_\bvec (\u)$ due to sampling with
$L_{\bvec}$  simply by first
subtracting each replica $X^{\bvec'}(\u -\v)$, $\v \in L^*_{\bvec}$, of every M-atom $\bvec'$ 
with $\|\bvec'\|>\|\bvec\|$, 
and then extracting the $\A^\bvec$
portion of the resulting ``de-aliased'' spectrum.

\vspace{1ex}

\begin{lemma}     \label{lem:nextHighestOmega} 
Suppose $M(B)$ is a Manhattan set and $x(\t)$ is an 
image whose spectrum $X(\u)$ is bandlimited to $\mathcal M(B)$.  
If  $X^{\bvec'}(\u)$ is known for all  $\bvec'$ larger than 
$\bvec$,  then    
%
%
%
%
\begin{align} 
    X^{\bvec}(\u) 
     &= H^\bvec(\u) \Big[ X_\bvec(\u) \, - \!\! 
	\sum_{\bvec' : \,  \| \bvec^\prime \| > \| \bvec \|}   \hspace{-3.5ex}
	 X^{\bvec^\prime}_\bvec (\u)  \Big]  \label{eq:nextHighestOmega} \, ,  
\end{align}
where $H^\bvec(\u)$ 
is defined in the previous lemma, 

\end{lemma}


\vspace{1ex}

\emph{Proof:} 
 \begin{align} 
     \text{RHS}\text{ of (\ref{eq:nextHighestOmega})} 
       &=   H^\bvec(\u) 
	    \Big[ \sum_{\v \in L^*_\bvec} \!\! X(\u-\v)  
	    - \!\!\!\!\! \sum_{\bvec' : \, \|\bvec^\prime\| > \|\bvec\|} 
	    \,\sum_{\v \in L^*_\bvec}
	    \!\! X^{\bvec^\prime}\!(\u-\v) \Big]  \label{eq:nextHighestOmega3} \\ 
         &=   H^\bvec(\u) 
	    \Big[  \sum_{\bvec' : \, \|\bvec^\prime\| \leq \|\bvec\|} 
	    \; \sum_{\v \in L^*_\bvec}
	    \! X^{\bvec^\prime}(\u-\v) \Big] \notag  \\ 
         &= X^{\bvec}(\u) \, ,  \notag
  \end{align} 
where the first equality uses (\ref{eq:sampledspectrum}), 
the second uses (\ref{eq:Xdecomposegamma}), and 
the last derives from Lemma \ref{lemma:SOmegaOverlap} and its corollary.
In particular, for the $\bvec' = \bvec$ term in the above sum, Part (a) of Lemma 
\ref{lemma:SOmegaOverlap}
applied with $\svec =  \bvec' = \bvec$ shows that 
 all replicas of
 $\A^\bvec$ (with $\v \neq \zerovec $) do not overlap $\A^\bvec$
 and, consequently, are eliminated by the filter $H^\bvec(\u)$.
Corollary \ref{cor:biggestGamma} implies every replica
of $\A^{\bvec'}$ (with $\v \neq \bf 0$) does not overlap $\A^\bvec$ and, consequently,
is again eliminated by the filter.  Since also $\A^{\bvec'}$ does not overlap $\A^{\bvec}$, 
the only term in the sum not eliminated by the filter is $X^\bvec(\u)$,
which establishes \eqref{eq:nextHighestOmega}.
%
 \hfill $\square$



\vspace{1ex}
Note that the sum in \eqref{eq:nextHighestOmega} can be limited
to $\bvec' \in \overline B$.  
\edits
%
Note also that Lemma \ref{lem:nextHighestOmega}  implies Lemma \ref{lem:highestOmega}, 
because when 
$\bvec$ is a largest weight bi-step vector in $\overline B$,
the summation term in \eqref{eq:nextHighestOmega} is zero, and so 
\eqref{eq:nextHighestOmega}   reduces to \eqref{eq:highestOmega}.
%
%
%

\iftoggle{comment}{\textbf{new stuff:   (new stuff looks good)}}{}
An alternative way to write \eqref{eq:nextHighestOmega} is 
\begin{align}
  X^\bvec(\u) &= H^\bvec(\u) \Big[ X_\bvec(\u) \, - \!\!\!\!\!
	\sum_{\bvec' : \,  \|\bvec^\prime\| > 	\|\bvec\|}   \,
	\sum_{\n \in C_\bvec} \!\! X^{\bvec^\prime}\!(\u-\n \odot \bm \beta_\bvec) \Big] 
  \label{eq:nextHighestOmega2}
\end{align}
where
$\bbeta_\bvec = (\beta_{\bvec,1}, \ldots, $ $\beta_{\bvec,d})$, with 
\begin{align*}
     \beta_{\bvec,i}  &~\triangleq~   \begin{cases}  
                {1 \over \lambda_i}, & \mbox{ if } b_i = 1 \\
                {1 \over k_i \lambda_i}, & \mbox{ if } b_i = 0 
                \end{cases}
\end{align*}
and 
\iftoggle{comment}{\footnote{temp footnote:  Matt's rationale for $C_\bvec$ definitions:  
If $b_i=1$, then only $n_i=0$ can overlap (since others are
contained in replicas of the fine Nyquist region $\Ncal_\onevec$).
if $b_i=0$, then everything between
sites at $\pm k_i$ can overlap, namely $|n_i|\leq k_i-1$}}{}
\begin{align*}
    C_\bvec  ~\triangleq~& \big\{ \n \in \mathbb Z^d : n_i =0 \mbox{ for $i$ s.t. $b_i=1$, and}
          | n_i | \leq k_i -1 \mbox{ for $i$ s.t. $b_i=0$} \big\}.
\end{align*}
To demonstrate \eqref{eq:nextHighestOmega2}, we note that 
since all atoms of the Manhattan partition are contained in $\Ncal_\onevec$, one can eliminate
from the last sum in \eqref{eq:nextHighestOmega3} any $\v$
such that $(\Ncal_\onevec + \v) \cap \Ncal_\onevec = \emptyset$.
This leads to limiting the sum to $\v$ such that $|v_i | < {1 \over \lambda_i}$ for each $i$.
Taking into account what $\v$'s are in $L^*_\bvec$ leads to 
\eqref{eq:nextHighestOmega2}.
For the usual 2D case, in which $B = \{ (1,0), (0,1) \}$, 
\eqref{eq:nextHighestOmega2} gives a different reconstruction formula
than in Section \ref{sec:twoD}    
for the spectrum in the coarse Nyquist region, $X^C(\u) = X^{(0,0)}(\u)$ 
(see \eqref{eq:SV2}-\eqref{eq:Xhatu}). Specifically, 
it subtracts terms involving both $X^V(\u) = X^{(0,1)}(\u)$ and $X^H(\u) = X^{(1,0)}(\u)$
from $X_C(\u)$, whereas the formula in Section \ref{sec:twoD} subtracts terms
involving $X^H(\u)$ 
\iftoggle{comment}{\footnote{\textbf{temp--} MP:  this could be either $X^H(\u)$ or $X_H(\u)$
depending on whether we want to emphasize that it depends on the original samples or
the recovered spectrum. Thoughts?  DN currently favors $X^H(\u)$, but this might change.
Leaving it as $X^H(\u)$ makes it easy to see the similarity to what just discussed for
the subtractions based on \eqref{eq:nextHighestOmega2}. }}{}
from $X_V(\u)$.  
Moreover, the summation over $\n$ in \eqref{eq:nextHighestOmega2}
sums over approximately twice as many values of $\v$.  This is because it
conservatively includes all $\v \in L^*_\bvec$ such that 
$\Ncal_1 + \v \cap \Ncal_1 \neq \emptyset$, whereas the formula in Section
\ref{sec:twoD} includes only $\v$'s such that $\Ncal_V + \v \cap \Ncal_C \neq \emptyset$.
If desired $C_\bvec$, in \eqref{eq:nextHighestOmega2} could be replaced
by a smaller set $C_{\bvec,\bvec'}$ that depends on $\bvec'$
as well as $\bvec$. 

\iftoggle{comment}{\textbf{end new stuff:}}{}


\vspace{1ex}
The basic idea behind following theorem, which is the main result of this section,
is that the process of finding  $X^\bvec(\u)$ for smaller and smaller weight
$\bvec$'s can continue until $X^\zerovec$, the spectrum in $\A^\zerovec (\u) =
\mathcal N_\zerovec$, is found, and all of $X(\u)$ is known.   As a result,
$x(\tvec)$ will also be known. 

\vspace{1ex}
\begin{theorem} \emph{Multidimensional Manhattan Sampling Theorem.}  
\label{thm:multidimenManhattanthm} 
Suppose we sample an 
image
  $x(\t)$ with Manhattan set   $M(B)$.  If
  the image spectrum $X(\u) $ is bandlimited 
 to $\mathcal M(B)$, then for each $\bvec \in \overline B$,
 $X^\bvec(\u)$ can be exactly recovered from the samples with 
 the following ``onion-peeling'' approach --- 
 apply Lemma \ref{lem:highestOmega} for the largest $\bvec$'s
 in $\overline B$, and then repeatedly apply 
 Lemma \ref{lem:nextHighestOmega} for the next largest $\bvec$'s.
 Then, $x(\t)$ can be exactly recovered from
  \begin{equation*} 
      x(\t) ~=~ \mathcal F^{-1} \bigg\{ \sum_{
	\bvec\in \overline B} X^\bvec(\u)  \bigg\}  \, .    \label{eq:recoveryOfF} 
\end{equation*}
  \label{thm:multiManSampThm} 
\end{theorem}

\vspace{1ex}
\emph{Proof:} 
From (\ref{eq:Xdecomposegamma}) and the bandlimitation of 
$X(\u)$, it is clear that $X(\u)$ can be recovered if
$X^\bvec(\u)$ is recovered for each $\bvec \in \overline B$.  
First, $X^\bvec (\u)$ can be recovered via Lemma  \ref{lem:highestOmega}
for the  largest $\bvec$'s in $\overline B$,
which correspond to the highest frequency M-atoms.
Next,  repeatedly applying Lemma \ref{lem:nextHighestOmega} 
enables one to recover $X^\bvec(\u)$ for the Nyquist
atoms corresponding to the largest of the remaining $\bvec$'s, 
until  $X^{\zerovec}(\u)$,  corresponding to the lowpass atom, is recovered.  
\hfill $\square$

\vspace{1ex}
While this theorem indicates a frequency domain reconstruction,
followed by an inverse transform, a direct spatial domain
reconstruction is also possible.  As we now delineate, this
involves reconstructing each $x^\bvec(\t)$, $\bvec \in \overline B$,
and then using 
\begin{align*}
       x(\t) ~=~  \sum_{\bvec \in \overline B}  x^\bvec(\t) \, .
\end{align*}
Taking the inverse transform of 
(\ref{eq:nextHighestOmega2}) yields
\begin{align*}
  x^\bvec(\t) &~=~  h^\bvec(\t) \star \Big[ x_\bvec(\t) 
           \,  -  \!\!\!\! \sum_{\|\bvec'\| > \|\bvec\|} \!\!\!\!\!  x_\bvec^{\bvec'} (\t) \Big] \\[.5ex]
      &~=~  h^\bvec(\t) \star \Big[ 
           K(\bvec) \!\! \sum_{\t' \in L_\bvec} \!\! \delta(\t -\t')  
           \Big( x(\t)    -  \!\!\!\! \sum_{\|\bvec'\| > \|\bvec\|} \!\!\!\!\!\! x^{\bvec'} (\t) \Big) \Big] \\[.5ex]
     &~=~    K_\bvec \sum_{\t' \in L_\bvec}   \!\!
           \Big( x(\t') \,   -  \! \! \sum_{\|\bvec'\| > \|\bvec\|} \!\!\! x^{\bvec'} (\t') \Big) 
            \,   h^\bvec  (\t-\t') \, ,
\end{align*}
where $h^\bvec(\t) = \mathcal F^{-1} \big\{ H^\bvec(\u) \big\}$,
 $\star$ denotes convolution, and $K_\bvec = \prod_{i : b_i = 1} \lambda_i  \times \prod_{i : b_i=0 }  k_i \lambda_i$.  \edits This shows how $x^\bvec(\t)$ can be
found --- first for the largest $\bvec$'s from samples of $x(\t)$ taken on $L_\bvec$,
then for the next largest $\bvec$'s from samples of $x(\t)$ taken on $L_\bvec$, as well as samples of $x^{\bvec'}(\t)$ taken on $L_\bvec$ for 
all larger $\bvec'$, and so on.  It remains to find a formula for $h^\bvec(\t)$.

To find a formula for $h^\bvec(\t)$, which is the inverse transform
of $H^\bvec(\u)$, which in turn has support $\A^\bvec$,
we begin by recalling that $\A^\bvec$ is the union of 
$2^{\|\bvec\|}$ orthotopes in frequency space.  
Along each dimension $i$, these orthotopes
are centered at zero if $b_i =0$, and at $\pm c_i$ if $b_i=1$,
where
\begin{equation}
  c_i  ~\triangleq~ \frac{1}{2}\Big( \frac{1}{2\lambda_i} + \frac{1}{2k_i\lambda_i}  \Big) \, .
 \nonumber  \label{eq:cgami}
\end{equation}
Additionally, along the $i$th dimension, these orthotopes have length
\begin{equation} 
   w_i(\bvec) ~\triangleq~ \begin{cases} \frac{1}{2\lambda_i} -
  \frac{1}{2k_i\lambda_i}, & b_i=1\\[.5ex] 
     \frac{1}{k_i\lambda_i}, & b_i=0.  \end{cases}  \, . \nonumber  \label{eq:wgami} 
\end{equation}
Using these quantities, we can write the filter $H^\bvec(\u)$ as
\begin{equation}
      H^\bvec(\u) = \bigg[\prod_{i=1}^d \rect\Big({u_i \over w_i(\bvec)} \Big)
    \bigg] \star \bigg[\prod_{i : b_i=0} \delta(u_i) \prod_{i : b_i = 1}
    \Big[ \delta(u_i - c_i) + \delta(u_i + c_i)\Big]\bigg] \, ,
     \nonumber  \label{eq:Hrewrite} 
\end{equation}
where  $\rect(x) \triangleq 1$ for $|x| < {1 \over 2}$, and $ 0$ otherwise.
Note that the first term is an orthotope centered at $\bm 0$, and the convolution with
delta functions shifts the orthotopes along all dimensions $i$ 
such that $b_i =1$.  
Taking the inverse transform yields 
\begin{equation} 
     h_\bvec(\t) \,=\,
  \prod_{i=1}^d  w_i(\bvec) \, \sinc ( w_i(\bvec) t_i)  
     \cdot
    \prod_{i : b_i = 1}  2 \cos(2\pi c_i t_i )  \,  ,  \nonumber  \label{eq:hspace} 
\end{equation}
where $ \sinc(t) \triangleq \frac{\sin \pi t}{\pi t}$.
Observe that, as mentioned in the introduction,  
these impulse responses depend on the $k_i$'s and $\lambda_i$'s,
but not the choice of bi-step lattices that comprise the Manhattan set.  Moreover,
the $\lambda_i$'s have only a simple spatial scaling effect on the filters.

\subsection{Achievement of Landau lower bound on sampling density}
\label{sec:landau}


We now show that the volume of $\mathcal M( B)$,
denoted $| \M (B) |$,
equals the sampling density of $M(B)$.  As a result,
the set of images bandlimited to $\mathcal M(B)$ is a maximal
set of images that are reconstructable from sample set $M(B)$.
Equivalently, $M(B)$ has the smallest density of any
sampling set such that all images bandlimited to $\mathcal M(B)$
are reconstructable.

Since the M-atoms
partition $\M(B)$, we can calculate $|\M(B)|$ simply by summing over
the volumes of the M-atoms $\A^\bvec$ for $\bvec \in \overline
B$:
\begin{align*}
  | \M(B)  |
  &= \sum_{\bvec \in \overline B} |\A^\bvec| 
  = \sum_{\bvec \in \overline B} \! \Big(\!\!
     \prod_{i: b_i=1} 
     \!\!\! 2 \Big( \frac{1}{2\lambda_i}-\frac{1}{2k_i\lambda_i} \Big)
     \!\!\! \prod_{i : b_i=0} 
     \!  \frac{1}{k_i\lambda_i}  \Big) \\[.5ex]
  &~=~ \frac{\sum_{\bvec\in \overline B} 
     \prod_{i : b_i=1}(k_i-1)}{\prod_{i=1}^d k_i\lambda_i}
  \label{eq:densityFreq}
\end{align*}
Comparing the above to (\ref{eq:rhoGamma}), we see that $| \M(B) |$ 
equals the sampling density $\rho(B)$.

\subsection{Discrete-space images}


\edits
The $d$-dimensional Manhattan sampling theorem and reconstruction 
procedures can be straightforwardly extended to  discrete-space images in $d$ dimensions in the
same fashion as for two dimensions.  For example, for infinite-support images,
frequencies  need to be scaled by $2\pi$, and for finite-support images, 
each spatial resolution $T_i$  must be a multiple of $k_i \lambda_i$ 
and the Manhattan atoms $\widetilde A_\bvec$  need to be redefined  to be consistent
with the DFT, as was done for the discrete Nyquist 
region $\widetilde \Ncal_{\alpha_1,\alpha_2}$.   
Here, we simply give the main step
of the frequency-space  onion-peeling reconstruction algorithm for reconstructing 
a Manhattan bandlimited discrete-space image $x[\t]$ with finite support sampled with Manhattan set $M(B)$:
\begin{align*}
    X{^\bvec} [\t] 
        &\,=\,  \widetilde H^\bvec [\u] \;     \text{DFT} \bigg\{ x_\bvec [\t]  
      \, - \hspace{-2ex} \sum_{\bvec' : \| \bvec' \| > \| \bvec \|}   
      \hspace{-3ex} x^{\bvec'}_\bvec[\t]  \bigg\}  \, ,
\end{align*}
where $x_b[\t]$ and $x^{\bvec'}_\bvec[\t]$ denote, respectively, the $L_\bvec$ 
subsamplings of the Manhattan samples (scaled by $K_\bvec$), and 
the previously reconstructed atom $x^{\bvec'}[\t]$, 
and $\widetilde H^\bvec[\u]$ denotes an ideal bandpass filter 
for atom $\widetilde \A_\bvec$.

\section{Concluding Remarks}
\label{sec:conclusions}


%




\iftoggle{comment}{{\bf (considerable modifications)}}{}
In two dimensions, this paper has  shown that from samples of a Manhattan set one can
perfectly reconstruct any image
that is bandlimited to the union of the Nyquist regions of the horizontal and vertical
rectangular lattices comprising the Manhattan set.  It also prescribed a straightforward
linear reconstruction procedure, for continuous- and discrete-space images.

For three and higher dimensions, this paper has identified Manhattan sets as the union
of a finite number of bi-step rectangular lattices, with the result that many 
Manhattan geometries are possible.  
It introduced an efficient binary-vector representation of bi-step lattices, and consequently
Manhattan sets, which enabled the specification of a partition of the dense rectangular lattice
into collections of cosets of the coarse lattice.  This, in turn, enabled the density of a Manhattan to be
computed.  The representation of bi-step lattices also enabled a partition of the Nyquist region of the dense
rectangular lattice, which in turn enabled a precise analysis of the aliasing, i.e., spectral overlaps,
by the atoms of any particular type in the spectral replicas induced by any particular bi-step lattice subsampling.  
With this, it was shown that images bandlimited to the union of the Nyquist regions of the bi-step lattices
comprising the Manhattan set can be perfectly reconstructed using an
an efficient closed-form onion-peeling type reconstruction algorithm that reconstructs
the image spectrum working from higher to lower frequency atoms of the partition.
At each step, the algorithm works with samples of one particular bi-step lattice (of the Manhattan
set), and obtains the spectrum of the image in the corresponding atom of the frequency
partition by subtracting  contributions due to aliasing of previously determined atoms 
of the partition. 
Both frequency- and time-domain versions of the algorithm were given.
It was also shown that the set of Manhattan bandlimited images is maximal in the Landau sense.
To the best of our knowledge, this is the first demonstration that images bandlimited
to the union of Nyquist regions can be recovered from the union of the corresponding
lattices.


There are several avenues for future research.  One could seek to extend
the results to continuous-space images whose spectra contain delta functions,
e.g. periodic images.  Second, instead of a recursive onion-peeling 
reconstruction, one could seek direct closed-form linear reconstructions,
as in \cite{faridani90}, which might be useful for implementations,
though they might have less intuitive appeal.   
This is not difficult in 2D, but is more challenging in 
higher dimensions. 
Finally, whereas M-sampling can be viewed
as sampling (in various ways) on the boundaries of a rectangular (hyper-rectangular) lattice tessellation, one could seek sampling theorems and
reconstruction procedures for images sampled on the boundaries of other
lattice tessellations, such as a hexagonal tessellation.
\bibliographystyle{IEEEtran}
%


\bibliography{multiManhattanSampJournal}

%








\end{document}